\documentclass[a4paper,11pt]{article}
\pdfoutput=1 

\usepackage{jcappub} 
\usepackage{lineno}

\usepackage[T1]{fontenc} 
\usepackage[utf8]{inputenc}
\usepackage{amsmath,amsfonts,amssymb,amsthm,mathtools,array}
\usepackage{xcolor}
\usepackage{tensor}
\usepackage{bm}
\usepackage{float}
\usepackage{hyperref}
\usepackage{enumitem}
\usepackage{tcolorbox}
\usepackage{comment}
\usepackage{multicol}
\usepackage{graphicx}
\usepackage{subfig}
\usepackage{caption}
\usepackage{aas_macros}
\newcommand{\mi}{\mathrm{i}}

\subheader{
\footnotesize
\vspace*{-3.7em}
\begin{flushright}
RBI-ThPhys-2023-06
\end{flushright}
}

\definecolor{Grn}{rgb}{0.,0.6,0.}

\title{Effective Field Theory of Intrinsic Alignments at One Loop Order: a Comparison to Dark Matter Simulations}

\author[a]{Thomas Bakx,}
\author[b,c]{Toshiki Kurita,}
\author[a]{Nora Elisa Chisari,}
\author[d,e,f]{Zvonimir~Vlah,}
\author[g]{and Fabian~Schmidt.}

\affiliation[a]{Institute for Theoretical Physics, Utrecht University, Princetonplein 5, 3584 CC, Utrecht, The Netherlands.}
\affiliation[b]{Kavli Institute for the Physics and Mathematics of the Universe (WPI),
The University of Tokyo Institutes for Advanced Study (UTIAS),
The University of Tokyo, Chiba 277-8583, Japan}
\affiliation[c]{Department of Physics, Graduate School of Science, The University of Tokyo, 7-3-1 Hongo, Bunkyo-ku, Tokyo 113-0033, Japan}
\affiliation[d]{Ru\dj er Bo\v{s}kovi\'c Institute, Bijeni\v{c}ka cesta 54, 10000 Zagreb, Croatia}
\affiliation[e]{Kavli Institute for Cosmology, University of Cambridge, Cambridge CB\
3 0HA, UK}
\affiliation[f]{Department of Applied Mathematics and Theoretical Physics, University of Cambridge, Cambridge CB3 0WA, UK }
\affiliation[g]{Max Planck Institute for Astrophysics, Karl-Schwarzschild-Str. 1, 85748 Garching, Germany}

\emailAdd{t.j.m.bakx@uu.nl}

\abstract{
We test the regime of validity of the effective field theory (EFT) of intrinsic alignments (IA) at the one-loop level by comparing with 3D halo shape statistics in N-body simulations.
This model is based on the effective field theory of large-scale structure (EFT of LSS) and thus a theoretically well-motivated extension of the familiar non-linear alignment (NLA) model and the tidal-alignment-tidal-torquing (TATT) model. It contains a total of $8$ free bias parameters. Specifically, we measure the dark matter halo shape-shape multipoles $P_{EE}^{(0)}(k), P_{EE}^{(2)}(k), P_{BB}^{(0)}(k), P_{BB}^{(2)}(k)$ as well as the matter-shape multipoles $P_{\delta E}^{(0)}(k), P_{\delta E}^{(2)}(k)$ from the simulations and perform a joint fit to determine the largest wavenumber $k_{\text{max}}$ up to which the theory predictions from the EFT of IA are consistent with the measurements. We find that the EFT of IA is able to describe intrinsic alignments of dark matter halos up to $k_\text{max}=0.30\,h$/Mpc at $z=0$. This demonstrates a clear improvement over other existing alignment models like NLA and TATT, which are only accurate up to $k_\text{max}=0.05\,h$/Mpc. We examine the posterior distributions of the higher-order bias parameters, and show that their inclusion is necessary to describe intrinsic alignments in the quasi-linear regime. Further, the EFT of IA is able to accurately describe the auto-spectrum of intrinsic alignment B-modes, in contrast to the other alignment models considered.} 

\begin{document}

\maketitle
\flushbottom
 
\section{Introduction}
Intrinsic alignments are correlations between galaxy shapes that originate in gravitational interactions throughout the large-scale structure \cite{Joachimi15,Troxel15,Kiessling15,Kirk15}. They have been detected in multiple photometric and spectroscopic surveys \cite{Brown02,Mandelbaum06,Hirata07,Joachimi11,Singh15,Johnston18,Fortuna21} to high significance. Intrinsic alignments are a well-known contaminant to weak gravitational lensing surveys that can lead to biased results for cosmological parameters \cite[e.g.][]{Joachimi10,Kirk12,Krause16}. This applies already to Stage-III surveys \cite{Hikage19,Heymans21,Secco22}, but will become more problematic with the decreasing error bars that Stage-IV surveys will provide in the mid-2020s. Therefore, it is mandatory to model intrinsic alignments as accurately as possible. If this is done within a well-motivated theoretical framework, it further allows for a physical interpretation of the intrinsic alignment constraints. 

Moreover, intrinsic alignments are emerging as a cosmological source of information, most recently through measurements of the growth rate in \cite{Okumuraf} and anisotropic primordial non-Gaussianity \cite{Schmidt15,Chisari16,Kurita23} from spectroscopic surveys. In the future, it is expected that alignments can also constrain the expansion of the Universe through baryon acoustic oscillations \cite{ChisariDvorkin,vanDompseler} and sensitivity to primordial gravitational waves and parity-violation scenarios \cite{ChisariBICEP,Biagetti}. In addition, the alignments of clusters of galaxies are now also regularly measured and could play a role in cosmological constraints \cite{vanUitert,Vedder}. In this context, better models for intrinsic alignments would allow for exploiting the constraining power of smaller scales more robustly.

Over the past decade, it has become clear that the effective field theory of large-scale structure (``EFT of LSS'') \cite{Baumann12,Carrasco12,Porto14,Senatore15,Senatore15b,Perko16,Lewandowski18,Pajer_2013,Baldauf_2021,Carrasco_2014,Mirbabayi_2015} is successful at describing biased tracers of the dark matter density field in the quasi-linear regime, both in real space and in redshift space. The applications of the EFT of LSS have so far mostly been restricted to scalar quantities, e.g. galaxy or halo number densities. In particular, the theory has been successfully applied to galaxy clustering data sets in recent works \cite{Zhang22,Damico22,Piga22}, allowing for robust cosmological parameter extraction. 

It is possible to extend the EFT of LSS treatment to any symmetric tensor field $S_{ij}(\mathbf{x})$, for example the halo shape field which we will introduce momentarily. In reference \cite{Vlah20}, the power spectrum of halo shapes $S_{ij}$ (which include the diagonal part)
\begin{equation}
    (2\pi)^3P_{ijkl}(\mathbf{k})\delta^D(\mathbf{k+k'}) := \langle S_{ij}(\mathbf{k})S_{kl}(\mathbf{k'})\rangle
\end{equation}
was computed at one-loop order in the EFT of LSS. Galaxy shapes follow the same expansion; all physical effects of galaxy formation are absorbed by the free bias parameters \cite{Vlah20}. In addition, to enable a direct comparison to data, shapes should be projected on the sky. The prescription for how to perform such projection was presented in \cite{Vlah21}. Galaxy shapes can then be described in a basis of $E$ and $B$-modes and estimated from the actual data \cite{Kurita_2022}, similarly to the polarization of the cosmic microwave background \cite{Zaldarriaga97}. These results will be used and discussed at length in this paper.  

Previous attempts at describing intrinsic alignments of galaxies include (i) the LA model \cite{Catelan01,Hirata04}, (ii) the non-linear alignment (NLA) model \cite{Bridle07}, (iii) the `TATT' (Tidal Alignment - Tidal Torquing) model, either with or without the inclusion of velocity shear (VS) \cite{Blazek19,Schmitz18} and (iv) the halo model \cite{Schneider10,Fortuna21b}. 

The LA and NLA models posit a linear (but in general redshift-dependent) relation between the galaxy intrinsic shape field $g_{ij}(\mathbf{x},\eta)$, the trace-free part of $S_{ij}$, and the gravitational tidal field $K_{ij}(\mathbf{x},\eta)$.
The TATT model considers a more general expansion of the intrinsic shape field including more gravitational operators than just the tidal field. 
However, the authors of \cite{Blazek19} already pointed out that the perturbative expansion on which the TATT model relies is incomplete, as (i) it only expands to second order in the density field and (ii) it does not include higher-derivative contributions. 
The EFT of intrinsic alignments (``EFT of IA'') overcomes these two difficulties, by consistently incorporating the fact that galaxy and halo formation happens on long time scales. This in particular includes the limiting cases of expanding shapes at the initial (``Lagrangian expansion'') and final times (``Eulerian expansion''). 
Importantly, the LA and NLA models do not predict the presence of intrinsic $B$-mode auto-correlations (besides a constant shot noise contribution), while non-trivial $B$-mode auto-correlations appear at one-loop order in two-point statistics in the TATT and EFT models. The presence of $B$-mode autocorrelations thus necessitates the use of either of these latter models in order to explain this feature. 

The halo model \cite{Schneider10} in turn offers a possibility to reach the nonlinear regime through the assumption that all galaxies live in spherically symmetric halos, and that they align in a particular way with respect to the center of the halo. This model seems to perform very well in the context of weak lensing contamination \cite{Fortuna21b}, though it necessarily makes more physical assumptions than perturbative models to reach nonlinear scales. For example, it relies on the assumption that galaxies live in spherically symmetric halos and align their major axes in specific (observationally-motivated) ways around the center of the halo. 

To mitigate the impact of IA modelling in cosmological constraints from weak lensing, a conservative strategy would be to cross-check the fidelity of the cosmological constraints when different models and corresponding scale-cuts are applied. 

Intrinsic alignments can also contaminate galaxy clustering measurements by inducing selection effects along the line-of-sight that originate in a correlation between the galaxy orientation and the underlying large-scale structure \cite{Hirata09,Desjacques_2018_rs}. Several works have already pointed out that this effect can be present in existing analyses up to $3\sigma$ \cite{Martens18,Obuljen20} (though see \cite{Singh21}). For upcoming surveys, alignments are expected to bias redshift-space distortions (RSDs) constraints at higher significance \cite{Zwetsloot}. The EFT of IA could offer a viable quasi-linear model for mitigation in this context (similarly to the approach adopted in \cite{Agarwal21}). Free bias parameters would absorb the uncertainties related to how galaxies respond to the orientation-dependent selection effect, but of course only up to the scales where the EFT is valid. 

To validate the EFT of IA as a viable candidate amongst alignment models and determine the range of scales where the EFT is applicable, this work aims to subject it to a rigorous comparison with halos in dark-matter-only simulations, along with other alignment models like NLA and TATT. 
We emphasize that small-scale effects of baryons, such as pressure and feedback, are consistently incorporated in the EFT framework, including the description of alignments, as long as one restricts to scales where the perturbative expansion is valid. Nevertheless, we relegate a comparison with full hydrodynamic simulations to future work.

On the other hand, the alignment signal of halos in simulations is easier to isolate than that of observed galaxies, both because we have perfect knowledge of the halos' location while galaxy distances based on photometric redshifts are uncertain, and because halos tend to align more strongly than galaxies \cite{Tenneti14,Velliscig15,Chisari17}. In this sense, our results regarding the scale of validity $k_{\text{max}}$ of each alignment model can be considered conservative. In the future, we would like to directly determine the impact of adopting the EFT of IA as a model for weak gravitational lensing and clustering contamination in the context of full likelihood inference, and to understand how well biases in cosmological parameters can be mitigated with this model. However, this is outside of the scope of our current work.

This paper is organised as follows. We first review the theory of intrinsic alignments at next-to-leading order and the relation to other existing alignment models in Section \ref{theory}. We then discuss the specifics of the simulation suite in Section \ref{sims}, and show results from fitting several different alignment models to the simulations in Section \ref{results}. Appendices provide explicit expressions for the perturbation theory kernels needed to 
compute the EFT of IA predictions and a justification for neglecting selection effects.

Throughout the entire paper, we use $c=1$. The fiducial cosmology we use for the theory computation and simulations is Euclidean $\Lambda$CDM with $\omega_b=0.02225, \omega_c=0.1198,\Omega_\Lambda = 0.6844, n_s=0.9645,\ln(10^{10}A_s)=3.094$ which is consistent with the {\it Planck} CMB data \cite{Planck2015}. The linear power spectrum was computed using CAMB \cite{camb}.

\section{Intrinsic Alignments at One Loop Order}\label{theory}

\subsection{Three-dimensional and projected galaxy shapes}

We can model the shape of any object by considering its inertia tensor $I_{ij}$. In this paper, we focus on dark matter halos in N-body simulations; however, the following discussion applies to any physical tracer. If it pertains to galaxies, one may consider a weighting of the expressions below by brightness, for example. 
If the dark matter halo consists of a cloud of points of equal mass at locations $\mathbf{x}_p$, then its inertia tensor is given by
\begin{equation}
I_{ij}\propto \sum_p w_p \Delta{\bf x}_{p,i}\Delta{\bf x}_{p,j}; \label{eq:Iij}
\end{equation}
where $w_p$ is some radial weight function and $\Delta{\bf x}_p$ is the difference between the coordinate of each point (e.g. dark matter particle) and the center of mass of the object. Different works adopt various choices of weight functions and in simulations, iterative updating schemes may also be employed to reduce noise \cite{Kurita_2020}. This choice is relevant in that it determines the signal-to-noise of the IA measurement, and leads to different values of the free parameters of the IA theory \cite{Singh16}. Note that $\text{Tr}(I_{ij})$ is a positive-definite quantity that can be used to define the size of a galaxy through 
\begin{equation}\label{size}
    \text{Tr}(I_{ij}) =: s^2.
\end{equation}
One can then write down a formal expression for the `inertia tensor field', i.e.  
\begin{equation}
    I_{ij}(\mathbf{x}):=\sum_\alpha I_{ij}({\bf x}_\alpha)\delta({\bf x}-{\bf x}_\alpha);
\end{equation}
where the sum is over all objects $\alpha$ and $\mathbf{x}_\alpha$ is the center of mass position of the object $\alpha$. Note that by statistical isotropy we must have $\langle I_{ij} \rangle \propto \delta_{ij}$, so by virtue of Eq. \eqref{size} we can write $\langle I_{ij} \rangle = \delta_{ij}\bar{s}^2/3$, where $\bar{s}^2 = \langle \text{Tr}(I_{ij}) \rangle$. The 3D shape field perturbation can then be defined as 
\begin{equation}\label{shapedef}
    S_{ij}({\bf x}) := \frac{I_{ij}({\bf x})-\langle I_{ij} \rangle}{\langle\text{Tr}(I_{ij})\rangle} = \frac{1}{3}\delta_{ij}\delta_s({\bf x}) + g_{ij}({\bf x}).
\end{equation}
The tracefree part $g_{ij} = \text{TF}(S)_{ij}$ will be referred to as the intrinsic shape field, while $\delta_s$ is the size perturbation field defined through $s^2(\mathbf{x}) =:  \bar{s}^2(1+\delta_s(\mathbf{x}))$. The intrinsic shape field transforms as a tracefree tensor under 3D rotations. Note that since the size is a scalar, $\delta_s$ can be expanded just like tracer number counts \cite{Vlah20}. Here we choose to normalize $S_{ij}$ by dividing a spatially constant factor, but other nonconstant normalizations of the shape field are also possible, \textit{as long as they transform as scalars under 3D rotation} and can be expanded in perturbation theory. These reinterpretations merely lead to a redefinition of the bias parameters in the expansions for $\delta_s$ and $g_{ij}$. 

Shapes are observed in projection on the sky. More precisely, we can take the line-of-sight to be the $x^3$ direction and define the projected intrinsic shape tensor \cite{Vlah21},  
\begin{equation}\label{elliptens}
\begin{aligned}
    \gamma_{ij,I}(\mathbf{x},z) &:= \text{TF}(\mathcal{P}^{ik}(\mathbf{\hat{n}})\mathcal{P}^{jl}(\mathbf{\hat{n}})g_{kl}(\mathbf{x},z)) 
    \\
    &= \frac{1}{2}\bigg( \mathcal{P}^{ik}(\mathbf{\hat{n}}) \mathcal{P}^{jl}(\mathbf{\hat{n}}) +\mathcal{P}^{il}(\mathbf{\hat{n}}) \mathcal{P}^{jk}(\mathbf{\hat{n}}) -\mathcal{P}^{ij}(\mathbf{\hat{n}}) \mathcal{P}^{kl}(\mathbf{\hat{n}}) \bigg)g_{kl}(\mathbf{x},z) \\
    &= \mathcal{P}^{ijkl}(\mathbf{\hat{n}})g_{kl}(\mathbf{x},z);
\end{aligned}
\end{equation}
where
\begin{equation}
    \mathcal{P}^{ij}(\mathbf{\hat{n}}):=\delta^{ij}-\mathbf{\hat{n}}^i\mathbf{\hat{n}}^j 
\end{equation}
is a projection operator in the $\mathbf{\hat{n}}$-direction and the last line of Eq. \eqref{elliptens} is the definition of the total projection tensor $\mathcal{P}_{ijkl}$. While $g_{kl}$ has five degrees of freedom, the projected field only has two (one Euler angle, one axis ratio). Using Eq. \eqref{shapedef} it is easy to see that 
\begin{equation}
    \gamma_1 := \gamma_{11,I}=-\gamma_{22,I} \propto I_{11}-I_{22}, \qquad \gamma_2 := \gamma_{12,I}=\gamma_{21,I} \propto 2I_{12}. 
\end{equation}
Here the proportionality refers to the arbitrary normalization of the shape field mentioned above (i.e. the constant of proportionality is the same for both $\gamma_1$ and $\gamma_2$). 
The projected ellipticity of the object $\alpha$ is typically defined as 
\begin{equation}\label{12}
\begin{aligned}
\hat{\gamma}_1({\bf x}_\alpha)&:= \frac{I_{11}({\bf x}_\alpha)-I_{22}({\bf x}_\alpha)}{I_{11}({\bf x}_\alpha)+I_{22}({\bf x}_\alpha)};\\
\hat{\gamma}_2({\bf x}_\alpha)&:= \frac{2I_{12}({\bf x}_\alpha)}{I_{11}({\bf x}_\alpha)+I_{22}({\bf x}_\alpha)}.
\end{aligned}
\end{equation}
A projected intrinsic shape field can then be constructed by summing over all the projected shapes of objects present in the cosmological volume and interpolating over the positions of the objects, i.e.\footnote{In \cite{Kurita_2020}, a different notation is adopted and $\hat{\gamma}_{(1,2)}({\bf x})$ corresponds to $\hat{\gamma}_{(+,\times)}({\bf x})$.}
\begin{equation}\label{gamma12}
\hat{\gamma}_{(1,2)}({\bf x}) := \frac{1}{\bar{n}}\sum_\alpha \hat{\gamma}_{(1,2)}({\bf x}_\alpha) \delta^D({\bf x}-{\bf x}_\alpha).
\end{equation}
Here, $\bar{n}$ is the mean density of objects. 
One of the coordinates of the three-dimensional inertia tensor is thus effectively removed and shapes are normalized using the `projected area' $I_{11}+I_{22}$. Crucially, this quantity is in fact \textit{not} a scalar under 3D rotations, and its expansion in terms of the dark matter density perturbation could in principle contain terms that depend on the line-of-sight direction $\mathbf{\hat{n}}$. The fact that the ellipticities are normalized by the projected area 
 on an object-by-object basis means that they now have a \textit{nonlinear} dependence on the tensorial projection operator $\mathcal{P}^{ij}\mathcal{P}^{kl}$, thus introducing additional angular dependencies. These have been essentially overlooked in previous approaches to intrinsic shape statistics. In Appendix~\ref{App:D_L_S} we argue that in the cases studied in this paper, we can proceed by ignoring these effects; however, we caution against neglecting such line-of-sight-dependent contributions in general. In other words, we will derive theoretical expressions for $\gamma_1,\gamma_2$ by assuming that $\gamma_{ij,I}$ is given by Eq. \eqref{elliptens} (where the 3D intrinsic shape field $g_{ij}$ is statistically isotropic), and compare these to the measured ellipticities from Eq. \eqref{12} without any additional modifications. 

 This discussion naturally raises the question of what estimator to use for ellipticities in practice. While the problem described above could be avoided by normalizing by the ensemble average $\langle I_{11} + I_{22} \rangle$ rather than on an object-by-object basis, this would be a more noisy estimator. Indeed, outlier objects with very large projected sizes would contribute significantly to the total signal. It may be possible to mitigate this by normalizing by a different quantity like total mass or luminosity, or a function thereof, if this is feasible in practice. We leave these considerations for future work. 

In practice, there is also often a multiplicative ``responsivity'' factor applied on $\hat{\gamma}_{(1,2)}$ \cite{bernstein}. This is used to be able to relate intrinsic shapes to cosmological weak lensing shears. 
Note that the expression in Eq. \eqref{gamma12} is the number-density weighted shape field, i.e. we are only sampling at the positions of the halos. The density weighting of the shape field does not impact the applicability of the theory (see Section \ref{bias}). In Section \ref{projection} we explicitly relate the two shear ellipticity components to $g_{ij}$ and $S_{ij}$.   

The components $\gamma_{(1,2)}({\bf x})$ can further be transformed into $E$- and $B$-modes analogous to the CMB polarization. The power spectra of such $E$- and $B$-modes of intrinsic shapes can be computed in a very similar way to the power spectra of biased scalar tracers of the dark matter density field in the EFT of LSS. We briefly review this formalism here. 

\subsection{Bias Expansion for Symmetric Tensors}\label{bias}
In the EFT of LSS, one considers the expansion of any biased tracer of the dark matter density perturbation $\delta$ in terms of local operators $\mathcal{O}$ (we will turn to higher derivative effects and stochasticity in a moment) constructed out of $\delta$. The complicated nonlinear formation process and evolution of the tracer population are not amenable to analytic treatments. However, at least on sufficiently large scales, the tracer density can still be described in terms of the dark matter density, at the expense of introducing a finite number of free parameters $b_{\mathcal{O}}(\eta)$ \cite{Desjacques_2018}. More precisely, we write\footnote{The subscript $s$ is rather arbitrary, i.e. $\delta_s$ could refer to the halo number density perturbation, but also to any other scalar tracer. The r.h.s. of the expansion always takes the same form and the precise nature of the tracer is encoded in the values of the bias parameters $b_\mathcal{O}$. A similar remark also applies to the traceless part of $S_{ij}$ and hence, after projection, to the fields $\gamma_1(\mathbf{x}),\gamma_2(\mathbf{x})$.}
\begin{equation}\label{trace}
    \delta_s(\mathbf{x},\eta) = \sum_\mathcal{O}b_\mathcal{O}(\eta)\mathcal{O}(\mathbf{x},\eta).
\end{equation}
Here $\mathcal{O}$ denotes a set of \textit{bias operators} and $b_\mathcal{O}$ are free bias parameters which can generically be functions of (conformal) time $\eta$. When employing such perturbative  expansions in calculating the correlators of $\delta_s$, one needs to perform a renormalization procedure \cite{McDonald:2006, Assassi:2014, Angulo:2015} that prescribes how the small scale contributions affect the values of bias parameters for a given coarse-graining scale $\Lambda$. 

For example, the first operator appearing in the expansion is simply $\delta$. Thus, to first order the expansion recovers the \textit{linear bias} relation\footnote{The superscript (n) indicates the $n$-th order part of a quantity in perturbation theory. Here $\delta^{(n)}$ is the n-th order term in the SPT expansion of $\delta$, see e.g. \cite{Bernardeau_2002}.} $\delta_h^{(1)} = b_1^s \delta^{(1)}$.
The operators $\mathcal{O}$ can be constructed order-by-order as follows \cite{Mirbabayi_2015,Desjacques_2018}: first, define 
\begin{equation}\label{pi1def}
   \Pi^{[1]}_{ij}(\mathbf{x},\eta) := \frac{\partial_i \partial_j}{\nabla^2} \delta(\mathbf{x},\eta) = K_{ij}(\mathbf{x},\eta) + \frac{1}{3}\delta_{ij}\delta(\mathbf{x},\eta);
\end{equation}
i.e. $K_{ij}$ is the gravitational tidal field. Then, construct recursively
\begin{equation}\label{pin}
    \Pi^{[n]}_{ij} (\mathbf{x},\eta):= \frac{1}{(n-1)!}\bigg((\mathcal{H}f)^{-1}\frac{D}{D\eta} \Pi^{[n-1]}_{ij}(\mathbf{x},\eta) - (n-1)\Pi^{[n-1]}_{ij}(\mathbf{x},\eta) \bigg).    
\end{equation}
Here $\frac{D}{D\eta} := \frac{\partial}{\partial \eta} + v^{i} \frac{\partial}{\partial x^i}$ while $v^i$ is the fluid velocity, $f := \frac{d \ln D_+}{d\ln a}$ the growth rate and $\mathcal{H}:=aH$ the conformal Hubble rate. One can prove that a complete set of (unrenormalized) operators that are at most order $n$ in $\delta$ is given by taking all scalar combinations of $\Pi^{[m]}_{ij}$, with $m=1,2,\dots n$. For example, at first order one only has\footnote{Here we drop the indices for convenience, and from now on we write $\text{Tr}(X^2) := X_{ij}X_{ji}, \text{Tr}(X^3) = X_{ij}X_{jk}X_{ki}$ etc., where the indices are summed over.} $\text{Tr}(\Pi^{[1]})=\delta$, while at second order one can form
\begin{equation}\label{secor}
    \bigg(\text{Tr}(\Pi^{[2]})\bigg)^{(2)}, \quad \bigg(\text{Tr}((\Pi^{[1]})^2)\bigg)^{(2)}, \quad \bigg((\text{Tr}(\Pi^{[1]}))^2 \bigg)^{(2)}.
\end{equation}
It is important to note that in general, the operator $\Pi^{[m]}$ contains contributions of order $m, m+1, \dots$, i.e. not just of order $m$, hence the superscript (2) in Eq. \eqref{secor}. For example, $\Pi^{[1]}$ contains the fully nonlinear density field $\delta$ and not just $\delta^{(1)}$. It would seem that when the bias expansion is truncated at second order, there are four bias parameters in total. However, the complete set given by Eqs. \eqref{pi1def}, \eqref{pin} is not independent at each order. There is one degeneracy between the operators in Eq. \eqref{secor} that takes the form\footnote{In fact, it continues to hold at any order that $\text{Tr}(\Pi^{[n]})$ is not independent and thus does not need to be included \cite{Desjacques_2018}.}
\begin{equation}\label{deg}
    7\bigg(\text{Tr}(\Pi^{[2]})\bigg)^{(2)} - 5\bigg(\text{Tr}((\Pi^{[1]})^2)\bigg)^{(2)} - 2\bigg((\text{Tr}(\Pi^{[1]}))^2 \bigg)^{(2)} = 0.
\end{equation}
Hence, there are only three independent bias parameters that enter at second order. Such linear degeneracies continue to appear at higher order amongst the scalar combinations of the $\Pi^{[m]}_{ij}$ and should thus be taken into account. Of course, one is free to pick any two independent operators at second order as a basis for the expansion, and different conventions lead to different definitions of bias parameters which are related through linear transformations. 

Analogously, one can also write down an expansion for a trace-free tensor $g_{ij}$: 
\begin{equation}\label{tracefree}
    g_{ij}(\mathbf{x},\eta) = \text{TF}(S)_{ij}(\mathbf{x},\eta) = \sum_\mathcal{O'}b_\mathcal{O'}(\eta)\mathcal{O}'_{ij}(\mathbf{x},\eta).
\end{equation}
Since $g_{ij}$ is symmetric and trace-free, all operators $\mathcal{O'}$ in the expansion must also exhibit these symmetries. In analogy to the scalar case, they can be constructed out of the operators $\Pi_{ij}^{[n]}$ by considering all \textit{trace-free} combinations. For example, at leading order there is only $\text{TF}(\Pi^{[1]}_{ij})^{(1)} = K_{ij}^{(1)}$, while at second order one now has 
\begin{equation}
    \bigg(\text{TF}(\Pi^{[2]})\bigg)^{(2)}, \quad \bigg(\text{TF}((\Pi^{[1]})^2)\bigg)^{(2)}, \quad \bigg(\text{Tr}(\Pi^{[1]})\text{TF}(\Pi^{[1]}) \bigg)^{(2)}.
\end{equation}
However, for the trace-free operators there exists no analogue of Eq. \eqref{deg}, thus there are three independent operators at second order. At the third order, however, degeneracies do appear. When computing the correlators of the $g_{ij}$ using this perturbative expansion, we are again required to perform renormalization of the bias coefficients. This procedure follows analogously to the scalar field case, and we refer the reader to the ref. \cite{Vlah20} for further details.

In addition to the local gravitational operators described above, a complete expansion of any scalar field in the EFT of LSS also requires \textit{nonlocal} terms which involve taking spatial derivatives of the dark matter density perturbation. At leading order in spatial derivatives, this amounts to expanding the dark matter density field as 
\begin{equation}
    \delta(\mathbf{k},\eta) = \delta^{(1)}(\mathbf{k},\eta)+\delta^{(2)}(\mathbf{k},\eta)+\delta^{(3)}(\mathbf{k},\eta) + b_{R}'(\eta)R^2 k^2 \delta^{(1)}(\mathbf{k},\eta);
\end{equation}
and treating the last term as third order in the matter density, where $b_R'$ is a free coefficient. Here the fixed scale $R$ can be thought of as the nonlinear scale. As such, Eq. \eqref{trace} picks up a term proportional to $\nabla^2 \delta^{(1)}(\mathbf{x})$ whereas Eq. \eqref{tracefree} gets an additional term $\propto \nabla^2 K_{ij}^{(1)}(\mathbf{x})$. 

Finally, in order to correctly describe the impact of small-scale (i.e. highly nonlinear) physics on a tracer of the dark matter field, one must introduce additional stochastic fields \cite{Desjacques_2018}. This implies that at leading order (which will be sufficient for our analysis) Eq. \eqref{trace} picks up a term $\epsilon(\mathbf{x},\eta)$ while Eq. \eqref{tracefree} picks up a term $\epsilon_{ij}(\mathbf{x},\eta)$ which is again trace-free\footnote{As argued in \cite{Vlah20}, stochastic terms of second and third order do not lead to any additional functionally independent contributions. We will neglect terms of order $k^2$ in the stochastic amplitudes as well.}. These stochastic fields are uncorrelated with any of the other operators $\Pi_{ij}^{[n]}$, and their two-point correlators take the form\footnote{From now on, a superscript `$s$' on a bias parameter or power spectrum refers to a scalar field while a superscript `$g$' refers to a traceless field, i.e. $g_{ij}$. Also, a prime superscript on a two-point correlator indicates that the factor of $(2\pi)^3$ and the Dirac delta $\delta^D(\mathbf{k+k'})$ have been omitted from the notation.}
\begin{align}
        \langle \epsilon(\mathbf{k})\epsilon(\mathbf{k'})\rangle' &= P_{\epsilon}^s (k); & P_{\epsilon}^s (k) &= c^s + \mathcal{O}(R^2k^2); \\ 
        \langle \epsilon_{ij}(\mathbf{k})\epsilon(\mathbf{k'})\rangle' &= (\mathbf{\hat{k}}_i\mathbf{\hat{k}}_j-\frac{1}{3}\delta_{ij})P_{\epsilon}^{gs} (k); & P_{\epsilon}^{gs}(k) &= \mathcal{O}(R^2k^2); \\  
        \langle \epsilon_{ij}(\mathbf{k})\epsilon_{kl}(\mathbf{k'})\rangle' &= (\delta_{ik}\delta_{jl}+\delta_{il}\delta_{jk}-\frac{2}{3}\delta_{ij}\delta_{kl})P_{\epsilon}^{g} (k); & P_{\epsilon}^g(k) &= c^g + \mathcal{O}(R^2k^2).  
\end{align}
Reference \cite{Vlah20} also lists subleading stochastic contributions for the shape perturbation, which become relevant in the analysis of the bispectrum already at tree-level. 
\subsection{Spherical Tensor Decomposition}
Any symmetric tensor field $S_{ij}(\mathbf{k})$ can be decomposed into a trace and trace-free part: 
\begin{equation}
    S_{ij}(\mathbf{k}) = \frac{1}{3}\delta_{ij}\text{Tr}(S)(\mathbf{k}) + \text{TF}(S)_{ij}(\mathbf{k}).
\end{equation}
The trace and trace-free parts are separately invariant under rotations. In $D=3$ dimensions, the trace-free part can be further decomposed by using spherical tensors $(\mathbf{Y}_2 ^{(m)})(\mathbf{\hat{k}})_{ij}$ with $m=-2,-1,0,1,2$ denoting the helicity \cite{sakurai_napolitano_2017}. They are given by 
\begin{equation}\label{sphten}
\begin{aligned}
    (\mathbf{Y}_2 ^{(0)})(\mathbf{\hat{k}})_{ij} &:= \sqrt{\frac{3}{2}}(\mathbf{\hat{k}}_i\mathbf{\hat{k}}_j -\frac{1}{3}\delta_{ij}); \\
    (\mathbf{Y}_2 ^{(\pm1)})(\mathbf{\hat{k}})_{ij} &:=\sqrt{\frac{1}{2}}(\mathbf{\hat{k}}_i \mathbf{e}^\pm_j + \mathbf{e}^\pm_i \mathbf{\hat{k}}_j); \\
    (\mathbf{Y}_2 ^{(\pm 2)}) (\mathbf{\hat{k}})_{ij}&:= \mathbf{e}_i ^\pm \mathbf{e}_j ^\pm;
\end{aligned}
\end{equation}
where $i,j=1,2,3$ and $\mathbf{e}^\pm := \mp \frac{1}{\sqrt{2}}(\mathbf{e}_1 \mp \mi\mathbf{e}_2)$. The unit vectors $\mathbf{e}_1$ and $\mathbf{e}_2$ are defined by
\begin{equation}\label{basisvec}
    \mathbf{e}_1 := \frac{\mathbf{\hat{k}}\times \mathbf{\hat{n}}}{|\mathbf{\hat{k}}\times \mathbf{\hat{n}}|}; \quad \mathbf{e}_2 := \mathbf{\hat{k}}\times \mathbf{e}_1;
\end{equation}
and $\mathbf{\hat{n}}$ is an arbitrary unit vector not (anti-)parallel to $\mathbf{\hat{k}}$. The spherical tensors are orthonormal, traceless and closed under conjugation:
\begin{equation}\label{yprops}
    (\mathbf{Y}_2 ^{(m)})(\mathbf{\hat{k}})_{ij}(\mathbf{Y}_2 ^{(m')})(\mathbf{\hat{k}})_{ij}=\delta_{mm'}; \quad \delta_{ij}(\mathbf{Y}_2 ^{(m)})(\mathbf{\hat{k}})_{ij}=0; \quad (\mathbf{Y}_2 ^{(m)})(\mathbf{\hat{k}})^*_{ij}=(-1)^m(\mathbf{Y}_2 ^{(-m)})(\mathbf{\hat{k}})_{ij};
\end{equation}
where summation over repeated indices is implied. Additionally, a spherical tensor transforms into a combination of spherical tensors when the vector $\mathbf{\hat{k}}$ is rotated\footnote{This is an equality between invariant tensors, i.e. we use the notation $(\mathbf{Y}_l ^{(q)})(\mathbf{\hat{k}}) = (\mathbf{Y}_l ^{(q)})(\mathbf{\hat{k}})_{ij} \mathbf{w}^i \otimes \mathbf{w}^j$ where $\mathbf{w}^1, \mathbf{w}^2, \mathbf{w}^3$ is an orthonormal basis for $\mathbb{R}^3$. The Wigner matrix $D_{mq}^l(R)$ is unitary, so this identity amounts to a rotation \textit{in the space of tensors}.}: 
\begin{equation}
    (\mathbf{Y}_l ^{(m)})(R\mathbf{\hat{k}}) = \sum_{q=-l}^l D^l_{mq}(R)(\mathbf{Y}_l ^{(q)})(\mathbf{\hat{k}}) \quad \text{for } R \in SO(3),
\end{equation}
where $D^l_{mq}(R)$ is a $(2l+1) \times (2l+1)$ Wigner matrix. The decomposition of the symmetric tensor field now reads\footnote{Here $\mathbf{Y}_{ij}^{(m)}(\mathbf{\hat{k}})$ is a shorthand for $(\mathbf{Y}_2 ^{(m)})(\mathbf{\hat{k}})_{ij}$ and round brackets in the subscript indicate a symmetrization (without a factor of $1/2$).}
\begin{equation}\label{decomp2}
    S_{ij}(\mathbf{k}) = \frac{1}{3}\delta_{ij}S_0 ^0(\mathbf{k})+ \sum_{m=-2}^2 S_2^{m}(\mathbf{k})\mathbf{Y}_{ij} ^{(m)}(\mathbf{\hat{k}});
\end{equation}
where $S_0 ^0(\mathbf{\hat{k}}):= \text{Tr}(S)(\mathbf{\hat{k}})$. 

For this work, it will suffice to consider only correlations of the same tracer, i.e. 
\begin{equation}
    \langle S_{ij} (\mathbf{k})S_{kl}(\mathbf{k'})\rangle := (2\pi)^3\delta^D(\mathbf{k+k'})P^{SS}_{ijkl}(\mathbf{k}).
\end{equation}
As usual, due to statistical homogeneity, the two-point correlator is proportional to $\delta^D(\mathbf{k+k'})$. However, due to the fact that $S_{ij}$ transforms as a tensor under rotations, i.e. $S_{ij}(R\mathbf{k}) = R_{ik}R_{jl} S_{kl}(\mathbf{k})$, the power spectrum still depends on the direction of $\mathbf{k}$. However, for the component fields $S_0^0, S_2^m$ one has
\begin{equation}\label{powspeccomp}
    \langle S_l^{(m)}(\mathbf{k})S_{l'}^{(m')}(\mathbf{k'})\rangle = (2\pi)^3 \delta_{mm'}\delta^D(\mathbf{k+k'})P_{ll'} ^{(m)}(k).   
\end{equation}
In other words, the auto-power spectrum $P_{ijkl}^{SS}$ contains 7 independent scalar contributions after imposing statistical homogeneity and isotropy, i.e. $P_{00}^{(0)}, P_{02}^{(0)} (=P_{20}^{(0)}), P_{22}^{(m)} (m=0, \pm 1, \pm 2)$. Moreover, we also have 
\begin{equation}
    P_{ll'} ^{(m)}(k)^* = P_{ll'} ^{(-m)}(k);
\end{equation}
by virtue of Eq. \eqref{yprops} and the reality of the Dirac delta: $\delta^D(\mathbf{k})^* = \delta^D(-\mathbf{k})$. 

It turns out that if one also imposes invariance under parity transformations, then
\begin{equation}\label{parinv}
    P_{ll'} ^{(m)}(k) = P_{ll'} ^{(-m)}(k);
\end{equation}
and in that case, there will be 5 instead of 7 independent scalar contributions \cite{Vlah20}. Throughout the rest of this paper, we will indeed assume that parity invariance holds. 

Using Eqs. \eqref{decomp2}, \eqref{powspeccomp} and \eqref{parinv} we conclude that
\begin{equation}\label{fullcomp}
    P^{SS}_{ijkl}(\mathbf{k}) = \frac{1}{9}\delta_{ij}\delta_{kl}P_{00}^{(0)}(k) + \frac{1}{3}\delta_{(ij}\mathbf{Y}_{kl)}^{(0)}P_{02}^{(0)}(k) + \mathbf{Y}_{ij}^{(0)}\mathbf{Y}_{kl}^{(0)}P_{22}^{(0)}(k) + \sum_{m=1}^2 (-1)^m \mathbf{Y}_{(ij}^{(-m)}\mathbf{Y}_{kl)}^{(m)}P_{22}^{(m)}(k);
\end{equation}
where the factor of $(-1)^m$ arises from the fact that $\mathbf{Y}_{ij}^{(m)}(-\mathbf{\hat{k}})=(-1)^m \mathbf{Y}_{ij}^{(-m)}(\mathbf{\hat{k}})$ and we omit the $(\mathbf{\hat{k}})$ argument of the spherical tensors to avoid clutter. The component power spectra can be extracted by appropriate contractions:
\begin{equation}
\begin{aligned}
    P_{00}^{(0)}(k) &= \delta_{ij}\delta_{kl}P^{SS}_{ijkl}(\mathbf{k}); \\
    P_{02}^{(0)}(k) &= \delta_{ij}\mathbf{Y}_{kl}^{(0)}P^{SS}_{ijkl}(\mathbf{k}); \\
    P_{22}^{(m)}(k) &= \mathbf{Y}_{ij}^{(m)}\mathbf{Y}_{kl}^{(m)}P^{SS}_{ijkl}(\mathbf{k}).
\end{aligned}
\end{equation}
Thus, to compute all contributions to the helicity power spectra at one-loop order, one should compute $P_{ijkl}^{SS}(\mathbf{k})$ to next to leading order by plugging in the expansion for $S_{ij}(\mathbf{k})$ in terms of the $\Pi_{ij}^{[n]} (n=1,2,3)$ and subsequently contract the result with the appropriate spherical tensors. This was already carried out in \cite{Vlah20} and we summarize the outcome here. Some equations from \cite{Vlah20} required corrections and we indicate this explicitly when relevant.

\subsection{Results for Three-Dimensional Intrinsic Alignment Power Spectra}\label{iaspec}

\begin{figure}
    \centering
    \includegraphics[width=0.7\textwidth]{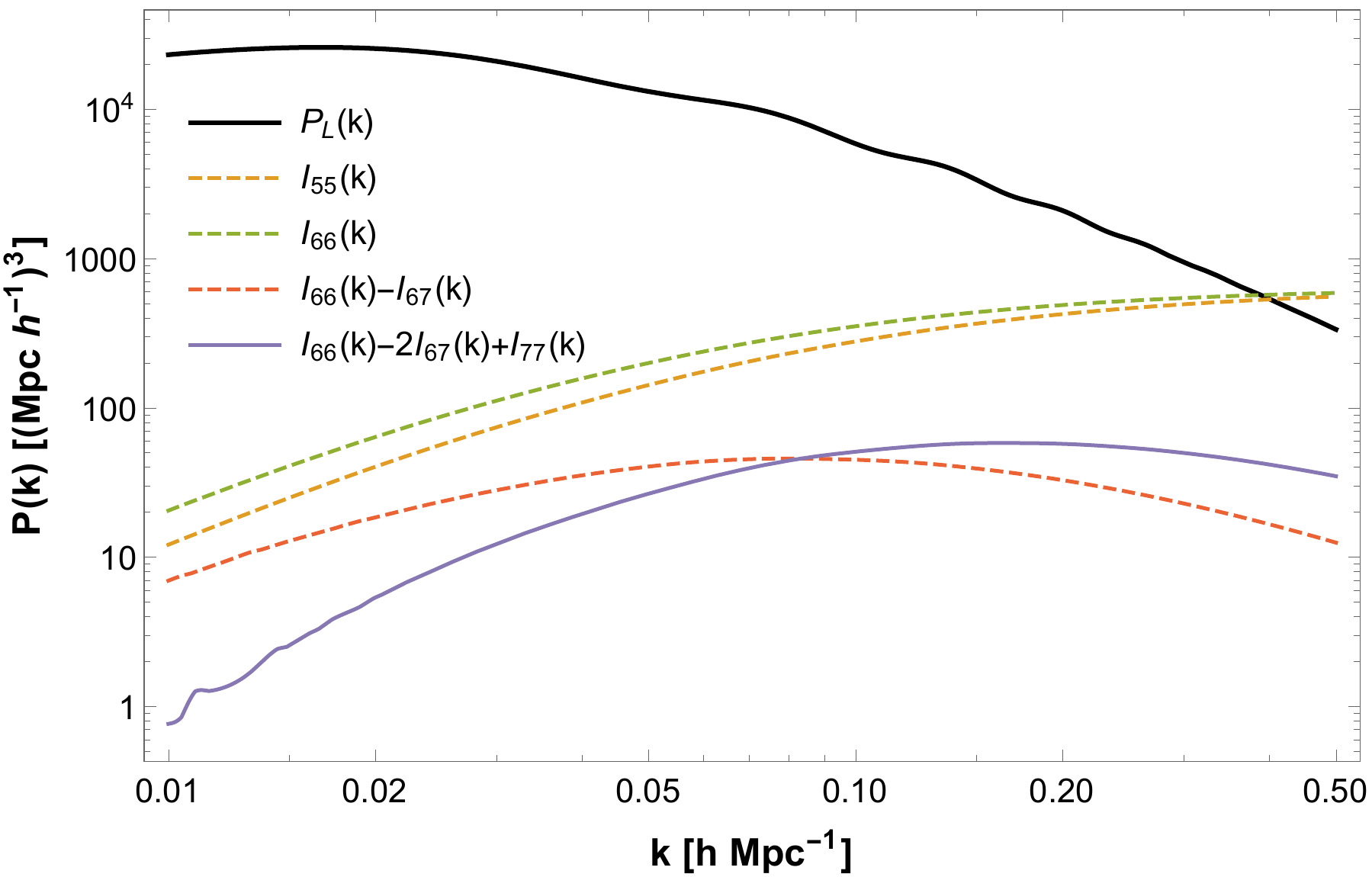}
    \caption{Some of the (22) contributions to the power spectra of intrinsic alignments at $z=0$. We show all contributions to $P_{BB}(k,\mu)$ for comparison (see Section \ref{projection}). The thick black curve represents the linear power spectrum. Dashed lines indicate negative contributions. The large-scale limit of each of the $I_{nm}$ has been subtracted.}
    \label{fig:iplot}
\end{figure}

\begin{figure}
    \centering
    \includegraphics[width=0.7\textwidth]{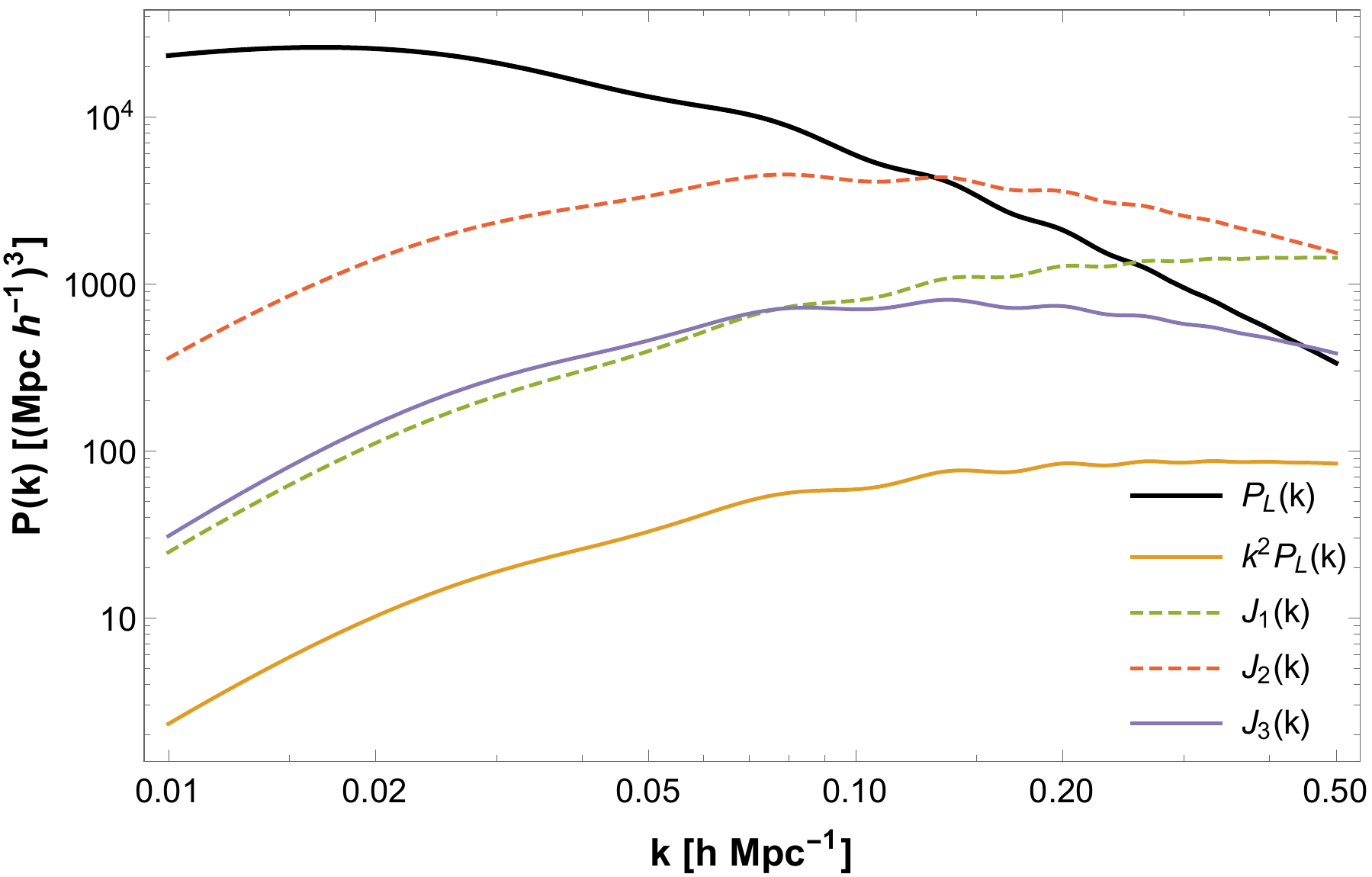}
    \caption{Contributions to the power spectra of intrinsic alignments at $z=0$ that are proportional to $P_L(k)$. The thick black curve represents the linear power spectrum. Dashed lines indicate negative contributions. }
    \label{fig:jplot}
\end{figure}

The results for the helicity power spectra can be separated into three pieces, namely (i) a part containing the linear and higher derivative contributions (L+H.D.), (ii) a (22)-part and a (13)+(31)-part coming from correlating two second order fields or a first and a third order field, respectively, and (iii) a stochastic contribution. Explicitly,
\begin{equation}\label{totspec}
    P_{ll'}^{(m)}(k) = [P_{ll'}^{(m)}]_{\text{L+H.D.}}(k)+[P_{ll'}^{(m)}]_{(22)}(k)+[P_{ll'}^{(m)}]_{(13)+(31)}(k)+[P_{ll'}^{(m)}]_{\epsilon}(k).
\end{equation}
We start with the linear and leading higher-derivative contributions. First, observe that
\begin{equation}\label{pi1}
   \Pi^{[1]}_{ij}(\mathbf{k}) = \frac{\mathbf{k}_i \mathbf{k}_j}{k^2} \delta(\mathbf{k}) = K_{ij}(\mathbf{k}) + \frac{1}{3}\delta_{ij}\delta(\mathbf{k}) = \left[\sqrt{2/3}\mathbf{Y}^{(0)}_{ij} + \frac{1}{3}\delta_{ij}\right]\delta(\mathbf{k}).
\end{equation}
Remembering that the (unbiased) dark matter density field can be expanded to third order as 
\begin{equation}
    \delta(\mathbf{k},\eta) = \delta^{(1)}(\mathbf{k},\eta)+\delta^{(2)}(\mathbf{k},\eta)+\delta^{(3)}(\mathbf{k},\eta) + b_{R}'(\eta)R^2 k^2 \delta^{(1)}(\mathbf{k},\eta);
\end{equation}
we can write the linear plus higher derivative contributions concisely as\footnote{Note again that we made use of the fact that the trace and trace-free parts of $\Pi^{[1]}_{ij}$ carry different bias coefficients. This corrects Eq. 4.9 from \cite{Vlah20}. It is implied that we neglect the two-loop $\mathcal{O}(k^4P_L(k))$ terms.}
\begin{equation}
    P^{\text{L+H.D.}}_{ijkl}(\mathbf{k})=\bigg(\frac{1}{3}\beta_L^s(k)\delta_{ij}+\beta_L^g(k) \sqrt{2/3}\mathbf{Y}^{(0)}_{ij}\bigg)\bigg(\frac{1}{3}\beta_L^s(k)\delta_{kl}+\beta_L^g(k) \sqrt{2/3}\mathbf{Y}^{(0)}_{kl}\bigg)P_L(k).
\end{equation}
We will set $R$ to $1$ $h^{-1}\,$Mpc for convenience from now on (thus rendering $b_R'$ dimensionless), but any change is of course reabsorbed into the definition of $b_R'$. The coefficients $\beta_L$ are defined as\footnote{Note that we redefined $b_R^{\{s,g\}} := b_R' b_1^{\{s,g\}} $.}
\begin{equation}
    \beta_L^{\{s,g\}} :=b_1^{\{s,g\}}+b^{\{s,g\}}_{R}R^2 k^2.
\end{equation} Thus, the linear and higher derivative contributions to the tracer auto-correlations are 
\begin{equation}\label{linpred}
\begin{aligned}
    \bigg[P_{00}^{(0)}\bigg]_{\text{L+H.D.}}(k)&=\bigg[(b_1 ^s)^2+2b_1^s b_{R}^s R^2k^2\bigg] P_L(k); \\
    \bigg[P_{02}^{(0)}\bigg]_{\text{L+H.D.}}(k)&=\sqrt{\frac{2}{3}}\bigg[b_1 ^s b_1^g+(b_1^s b_{R}^g+b_1^g b_{R}^s) R^2k^2\bigg]P_L(k); \\
    \bigg[P_{22}^{(0)}\bigg]_{\text{L+H.D.}}(k)&=\frac{2}{3}\bigg[(b_1 ^g)^2+2b_1^g b_{R}^g R^2k^2\bigg]P_L(k).
\end{aligned}
\end{equation}
Next, we move on to the next-to-leading order contributions. They can be split up into a (22)-contribution coming from correlating two fields that are second order in $\delta$ and a (13)+(31)-contribution coming from correlating a first and a third order field. The (22)-contributions can be expressed in terms of integrals $I_{nm}$ with $n,m=1,2,\dots 7, n\leq m$ that take the form 
\begin{equation}\label{inm}
    I_{nm}(k) = \int \frac{\mathrm{d}^3\mathbf{p}}{(2\pi)^3} K^I_{nm}(\mathbf{p},\mathbf{k-p})P_L(\mathbf{p})P_L(\mathbf{k-p});
\end{equation}
while the (13)+(31)-contributions are given in terms of integrals $J_n$ with $n=1,2,3$ that take the form
\begin{equation}\label{jn}
    J_n(k) = P_L(k) \int \frac{\mathrm{d}^3\mathbf{p}}{(2\pi)^3} K^J_n(\mathbf{k},\mathbf{p})P_L(\mathbf{p}).
\end{equation}
Explicit expressions for the perturbation theory kernels $K^I_{nm}, K^J_n$ are given in Appendix \ref{kernels}. Plots of some of the contributions are given in Figures \ref{fig:iplot}, \ref{fig:jplot}.
Omitting the wavenumber argument $k$ for brevity, the next-to-leading order parts of the intrinsic alignment power spectra $P_{ll'}^{(m)}$ are given by\footnote{This corrects Eqs. (5.5), (5.6) in \cite{Vlah20}. Note that the higher helicity power spectra $P_{22}^{(m)}\,(m=1,2)$ do not have $(13)+(31)$ contributions, because the lowest order contribution to the intrinsic shape field $g_{ij}$ is proportional to the gravitational tidal field $K_{ij}$, which is orthogonal to $\mathbf{Y}_2^{(m)}$ for $m=1,2$ (see Eq. \eqref{pi1}).}
\begin{equation}\label{eftpowspec}
\begin{aligned}
     [P_{00}^{(0)}]_{(22)} &= 2(b_1^s)^2 I_{11} + 4(b_1^s b_{2,1}^s)I_{12} + 2(b_{2,1}^s)^2 I_{22} + 4 (b_1^s b_{2,2}^s) I_{13} + 4(b_{2,1}^s b_{2,2}^s)I_{23} + 2(b_{2,2}^s)^2 I_{33}; \\
     [P_{00}^{(0)}]_{(13)+(31)} &=2(b_1^s)^2 J_{1} + 2(b_1^s b_{3,1}^s)J_2; \\
     [P_{02}^{(0)}]_{(22)} &= 2\sqrt{\frac{2}{3}}(b_1^s b_1^g)I_{11} + 2\sqrt{\frac{2}{3}}(b_{2,1}^s b_1 ^g)I_{12} + 2\sqrt{\frac{2}{3}}(b_{2,2}^s b_1 ^g)I_{13} + 2(b_1^s b_{2,1}^g)I_{14} + 2(b_{2,1}^s b_{2,1}^g)I_{24} \\
     &+2 (b_{2,2}^s b_{2,1}^g) I_{34} + \sqrt{\frac{1}{6}}(b_1 ^s b_{2,2}^g)(I_{13}-I_{12}) + \sqrt{\frac{1}{6}}(b_{2,1}^s b_{2,2}^g)(I_{23}-I_{22}) \\
     &+\sqrt{\frac{1}{6}}(b_{2,2}^s b_{2,2}^g )(I_{33}-I_{23});\\
     [P_{02}^{(0)}]_{(13)+(31)} &= 2\sqrt{\frac{2}{3}}(b_1^s b_1^g)J_1 + \sqrt{\frac{2}{3}}(b_{3,1}^s b_1 ^g + b_1 ^s b_{3,1}^g)J_2 + \sqrt{\frac{2}{3}}(b_1 ^s b_{3,2}^g)J_3;\\
     [P_{22}^{(0)}]_{(22)} &= \frac{4}{3}(b_1^g)^2I_{11}+4\sqrt{\frac{2}{3}}(b_1 ^g b_{2,1}^g)I_{14} + \frac{2}{3}(b_1 ^g b_{2,2}^g)(I_{13}-I_{12}) \\
     &+ \frac{1}{12}(b_{2,2}^g)^2(I_{22}-2I_{23}+I_{33})+ \sqrt{\frac{2}{3}}(b_{2,1}^g b_{2,2}^g)(I_{34}-I_{24})+ 2(b_{2,1}^g)^2I_{44};\\
     [P_{22}^{(0)}]_{(13)+(31)} &= \frac{4}{3}(b_1 ^g)^2 J_1+  \frac{4}{3}(b_1^g b_{3,1}^g)J_2 + \frac{4}{3}(b_1^g b_{3,2}^g)J_3;\\
     [P_{22}^{(1)}]_{(22)} &= 2(b_{2,1}^g)^2I_{55};\\
     [P_{22}^{(1)}]_{(13)+(31)} &= 0; \\
     [P_{22}^{(2)}]_{(22)} &= 2(b_{2,1}^g)^2I_{66}+4(b_{2,1}^g b_{2,3}^g)(I_{67}-I_{66})+2(b_{2,3}^g)^2(I_{66}-2I_{67}+I_{77}); \\
     [P_{22}^{(2)}]_{(13)+(31)} &= 0. 
\end{aligned}
\end{equation}


Again, we stress that these expressions are valid in the case of tracer auto-correlations, but they can straightforwardly be generalized. 
The leading plus next-to-leading order deterministic parts depend on $10$ bias parameters in total: 
\begin{equation}
    \underbrace{b_1^{\{s,g\}}}_{\text{linear}},\quad \underbrace{b_{2,1}^{\{s,g\}},b_{2,2}^{\{s,g\}},b_{2,3}^g}_{\text{second order}},\quad \underbrace{b_{3,1}^{\{s,g\}},b_{3,2}^g}_{\text{third order}}.
\end{equation}
Thus, the intrinsic shape field $g_{ij}$ requires two additional bias parameters $b_{2,3}^g$ and $b_{3,2}^g$ compared to the scalar biased tracer case. 

Finally, we discuss stochastic contributions. By contracting the stochastic power spectra with appropriate spherical tensors one has to leading order in stochasticity (i.e. neglecting $\mathcal{O}(k^2)$ terms)
\begin{equation}\label{stoch2}
\begin{aligned}
    [P_{00}^{(0)}]_{\epsilon}(k) &= 2c^s; \\
    [P_{02}^{(0)}]_{\epsilon}(k) &= 0; \\
    [P_{22}^{(m)}]_{\epsilon}(k) &= 2c^g; \qquad \text{for } m=0,1,2, \\
\end{aligned}
\end{equation}
where $c^s, c^g$ are (in general different) constants. In addition, some of the integrals $I_{nm}(k)$ asymptote to a nonzero constant as $k \to 0$, and this constant can always be reabsorbed by the stochastic contribution. Hence, we can subtract the $k \to 0$ limit from each $I_{nm}$ before fitting the model to data\footnote{Indeed, one can check by hand that each of the one-loop helicity spectra $P_{22}^{(m)}(k)$ in Eq. \eqref{eftpowspec} has the same $k \to 0$ limit, namely $\lim_{k\to 0}P_{22}^{(m)}(k) = \frac{2}{15\pi^2}(b_{2,1}^g)^2 \int p^2 P_L(p)^2 \mathrm{d}p.$ }. In doing this, we avoid $b_{2,1}^{g}$ being correlated with $c^{g}$ on very large scales. 

Thus, by plugging Eqs. \eqref{linpred}, \eqref{eftpowspec}, \eqref{stoch2} into Eq. \eqref{totspec} we obtain the full functional form of our model of three-dimensional intrinsic alignments. They depend on a total of $14$ free parameters, namely, 
\begin{equation}
    \underbrace{b_1^{\{s,g\}}}_{\text{linear}},\quad \underbrace{b_R ^{\{s,g\}}}_{\text{H.D.}},\quad \underbrace{b_{2,1}^{\{s,g\}},b_{2,2}^{\{s,g\}},b_{2,3}^g}_{\text{second order}},\quad \underbrace{b_{3,1}^{\{s,g\}},b_{3,2}^g}_{\text{third order}},\quad \underbrace{c^{\{s,g\}}}_{\text{stochastic}}.
\end{equation}
Note that all parameters are dimensionless, except for $c^{s,g}$ which has units$\text{ (Mpc}/h)^3$.
\subsection{Comparison to other Alignment Models}\label{models}

Several alignment models that have been discussed in the literature can be seen as special cases of the EFT of IA. We describe their connection to EFT of IA here explicitly.
The TATT (tidal alignment - tidal torquing) model from \cite{Blazek19} relies on the expansion 
\begin{equation}\label{tatt}
    g_{ij}(\mathbf{x}) = b_1^g K_{ij}(\mathbf{x})+ b_{\delta K} \delta K_{ij}(\mathbf{x}) + b_{KK}\text{TF}(K^2)_{ij}(\mathbf{x}).
\end{equation}
The first term corresponds to linear tidal alignment, while the second encodes density weighting of the tidal field and the third is the tidal torquing term. Since $\text{Tr}(\Pi^{[1]})\text{TF}(\Pi^{[1]})_{ij} = \delta K_{ij}$ and $\text{TF}((\Pi^{[1]})^2)_{ij} = \text{TF}(K^2)_{ij} + \frac{2}{3}\delta K_{ij}$ it becomes clear that the TATT model expansion for $g_{ij}$ is equivalent to the EFT expansion with the coefficient of $\text{TF}(\Pi^{[2]})_{ij}$ set to zero, as well as those of all third order operators in the shape expansion. The operator $\text{TF}(\Pi^{[2]})_{ij}$ contains the term 
\begin{equation}
t_{ij} =  -\bigg( \frac{\partial_i \partial_j}{\nabla^2}-\frac{1}{3}\delta_{ij}\bigg)(\delta + \frac{\theta}{\mathcal{H}f} );
\end{equation}
where $\theta = \nabla \cdot \mathbf{v}$, which thus contains the \textit{velocity shear} and is related to $\Pi^{[2]}$ via \cite{Desjacques_2018}
\begin{equation}
    \text{TF}(\Pi_{ij}^{[2]})^{(2)}=-\frac{5}{2}t_{ij}^{(2)}+\text{TF}((\Pi_{ij}^{[1]})^2)^{(2)}.
\end{equation}
This operator is also considered in the follow-up paper \cite{Schmitz18} of \cite{Blazek19}. In the notation of \cite{Vlah20}, the TATT model amounts to considering the second order expansion in their Eq. (B.37) with the constraint $c_{2,1}^g=0$. Thus, by following the arguments on pp. 55-58 there one arrives at a constraint among the bias parameters $b_{2,1}^g,b_{2,2}^g, b_{2,3}^g$ found by plugging in $c_{2,1}^g = 0$ in their Eq. (B.48). This yields $b_{2,2}^g = b_{2,3}^g$. In summary, we recover the expansion Eq. \eqref{tatt} by imposing $b_{2,2}^g = b_{2,3}^g$ and $b_{3,1}^g = b_{3,2}^g = b_R^g = 0$. 

If one includes the velocity shear term $t_{ij}$ in Eq. \eqref{tatt}, then the complete EFT of IA is recovered at second order. We will call this the `TATT+VS' model. It simply requires putting $b_{3,1}^g = b_{3,2}^g = b_R^g = 0$ with no further restrictions. The relevant basis transformation at second order is given by 
\begin{equation}\label{transf}
    \begin{pmatrix}
    b_{2,1}^g \\ b_{2,2}^g \\ b_{2,3}^g
    \end{pmatrix}
    =
    \begin{pmatrix}
        0 & 1/3 & 1 \\
        -8/7 & -2/3 & 1 \\
        0 & -2/3 & 1
    \end{pmatrix}
    \begin{pmatrix}
        b_t \\ b_{KK} \\ b_{\delta K}
    \end{pmatrix}.
\end{equation}
It is also possible to consider the galaxy bias expansion in Lagrangian space (fluid coordinates) rather than Eulerian space \cite{Schmitz18, Taruya_2021}. Specifically, it is shown in these works that at second order the Eulerian and Lagrangian bias parameters for the unweighted intrinsic shape field expansion $\bar{b}_t, \bar{b}_{KK}, \bar{b}_{\delta K}$ are related via
\begin{equation}
     \bar{b}_t^E = \bar{b}_t ^L + \frac{5}{2}\bar{b}_K^E,  \quad \bar{b}_{KK}^E = \bar{b}_{KK}^L - \bar{b}_K^E, \quad \bar{b}_{\delta K}^E = \bar{b}_{\delta K}^L - \frac{2}{3}\bar{b}_K^E.
\end{equation} 
where an overbar refers to the unweighted shape field (without density weighting). In the \textit{Lagrangian linear alignment model} (`Eulerian higher order bias' in the terminology of \cite{Taruya_2021}), it is assumed that all second order Lagrangian bias parameters $\bar{b}^L$ vanish. Under these conditions, by using $b_1^g = \bar{b}_K^E, b_t = \bar{b}_t ^E, b_{KK} = \bar{b}_{KK}^E, b_{\delta K} = b_1^s b_1^g + \bar{b}_{\delta K} ^E $ (where $b_1^s$ is the linear bias for the halo number density perturbation) we obtain the relations
\begin{equation}\label{lagr}
     b_{2,1}^g = (b_1^s-1)b_1^g, \quad b_{2,2}^g = (b_1^s-20/7)b_1^g, \quad b_{2,3}^g = b_1^s b_1^g.
\end{equation}
The linear Eulerian bias coefficient $b_K$ is negative in our case, which corresponds to tangential alignment of halos around overdensities. Moreover, for this sample of halos we have $b_1^s = 0.921 \pm 0.024$ which was obtained from fitting a linear bias model $\delta_{\rm h} = b_1^s \delta$ to $P_{\delta \rm h}$ on scales $k<0.05\,h$/Mpc. Therefore, provided that the Lagrangian linear alignment model holds at least approximately (this corresponds to the statement that the alignment of the halos is determined in the far past, where the dark matter density contrast behaves linearly), we see that $b_{2,2}^g$ is positive while $b_{2,3}^g$ is negative. We will use this fact as a prior on these bias parameters later and refer to it as the \textit{Lagrangian prior}. Note that the assumption of the TATT model, i.e. $b_{2,2}^g = b_{2,3}^g$, is inconsistent with the Lagrangian prior. Since $b_1^s \sim 1$, we decide not to put any prior on $b_{2,1}^g$. 

It is worth emphasizing that the DES Y3 analysis \cite{Secco22}  as well as the original TATT paper \cite{Blazek19} uses the \textit{fully nonlinear} matter power spectrum and hence is not identical to the analysis here. Our `TATT model' analysis amounts to expanding the matter density field to third order (thus keeping the speed of sound $b_R^s$) and the shape field to second order, while the original TATT model expands the matter density field to all orders.

Lastly, motivated by Figure \ref{fig:jplot} we decided to also examine a variant of the EFT of IA which drops two of the third order contributions $J_2, J_3$ by setting $b_{3,1}^g = b_{3,2}^g = 0$. Indeed, due to the similar shapes of $k^2P_L(k), J_2(k)$ and $J_3(k)$, and the fact that the helicity power spectra only depend on the combination $b_R^g k^2P_L(k) + b_{3,1}^g J_2(k) + b_{3,2}^g J_3(k)$, these contributions are expected to be highly degenerate on scales $k \lesssim 0.2 \,h$/Mpc anyway. This 6-parameter model is expected to yield similar results to the full EFT with fewer free parameters\footnote{Either way, since the third order parameters occur linearly in all the model predictions, marginalization over them can be performed analytically \cite{Philcox_2021} and hence does not yield any appreciable computational cost. Since we are interested in the posterior distributions of all bias parameters, we do not pursue this approach here. }.  

\subsection{Power Spectra of Projected Shape Fields}\label{projection}
Having obtained the three-dimensional predictions for the intrinsic alignment power spectra, we compute the spectra of projected shape variables. We adopt a specific line of sight $\mathbf{\hat{n}}$ (which we will take to coincide with the $z$-axis) along which a population of tracers is observed. We make use of the flat-sky (or plane-parallel) approximation, in which the region of the sky that is being observed is approximately planar and perpendicular to the line of sight (see \cite{Kurita_2022,Vlah21} for similar treatments). We can thus treat the line of sight as a fixed vector. The projected intrinsic shape field is then given by Eq. \eqref{elliptens}.

By construction, $\gamma_{ij,I}$ is a symmetric traceless tensor field. It follows that it can be decomposed by using the harmonic basis we introduced before. In fact, it is especially convenient to consider a harmonic basis $\mathbf{M}_{ij}^{(s)}$ in real space with respect to the line of sight, i.e.
\begin{equation}
\begin{aligned}
    (\mathbf{M}_2 ^{(0)})(\mathbf{\hat{n}})_{ij} &:= \sqrt{\frac{3}{2}}(\mathbf{\hat{n}}_i\mathbf{\hat{n}}_j -\frac{1}{3}\delta_{ij});\\
    (\mathbf{M}_2 ^{(\pm1)})(\mathbf{\hat{n}})_{ij} &:=\sqrt{\frac{1}{2}}(\mathbf{\hat{n}}_i \mathbf{m}^\pm_j + \mathbf{m}^\pm_i \mathbf{\hat{n}}_j); \\
    (\mathbf{M}_2 ^{(\pm 2)}) (\mathbf{\hat{n}})_{ij} &:= \mathbf{m}_i ^\pm \mathbf{m}_j ^\pm;
\end{aligned}
\end{equation}
where\footnote{One does not have to make this choice for $\mathbf{m}_1, \mathbf{m}_2$; any right-handed orthonormal basis $\mathbf{m}_1,\mathbf{m}_2,\mathbf{\hat{n}}$ suffices.} $\mathbf{m}_1 = \mathbf{\hat{x}}$, $\mathbf{m}_2 = \mathbf{\hat{y}}$ and $\mathbf{m}^\pm := \mp \frac{1}{\sqrt{2}}(\mathbf{m}_1 \mp \mi\mathbf{m}_2)$. Then, \begin{equation}
    \mathcal{P}^{ij}(\mathbf{\hat{n}})\mathbf{m}^{\pm}_j = \mathbf{m}^{\pm}_i.
\end{equation}
This is because $\mathbf{m}^+, \mathbf{m}^-$ are perpendicular to the line of sight. Since the projection operator vanishes in the $\mathbf{\hat{n}}$ direction, it follows using the same shorthand notation as before that 
\begin{equation}
\begin{aligned}
    \mathbf{M}_{ij}^{(\pm2)*}(\mathbf{\hat{n}})\mathcal{P}^{ijkl}(\mathbf{\hat{n}})  &= \mathbf{M}_{kl}^{(\pm2)*}(\mathbf{\hat{n}}); \\
    \mathbf{M}_{ij}^{(s)*}(\mathbf{\hat{n}})\mathcal{P}^{ijkl}(\mathbf{\hat{n}})  &= 0 \text{ for } s= -1,0,1. 
\end{aligned}
\end{equation}
Hence the degrees of freedom in $\gamma_{ij,I}$ are given by the $s=\pm 2$ helicity components $\gamma_{\pm 2}(\mathbf{x},z)$:
\begin{equation}
    \gamma_{I,ij}(\mathbf{x},z) =: \mathbf{M}_{ij}^{(+2)}(\mathbf{\hat{n}})\gamma_{+2}(\mathbf{x},z) +\mathbf{M}_{ij}^{(-2)}(\mathbf{\hat{n}})\gamma_{-2}(\mathbf{x},z).  
\end{equation}
Defining for the moment $\gamma_0(\mathbf{x},z):=\delta(\mathbf{x},z)$, we see that the components $\gamma_{s}(\mathbf{x},z)$ ($s=0,\pm 2$) are related to the 3D shape field perturbation $S_{ij}$ via  
\begin{equation}
    \gamma_{s}(\mathbf{x},z) = \mathbf{\bar{M}}_{ij}^{(s)*}(\mathbf{\hat{n}}) S_{ij}(\mathbf{x},z);
\end{equation}
where 
\begin{equation}
    \mathbf{\bar{M}}_{ij}^{(0)}(\mathbf{\hat{n}}) = \delta_{ij}; \qquad \mathbf{\bar{M}}_{ij}^{(s)}(\mathbf{\hat{n}}) = \mathbf{M}_{ij}^{(s)}(\mathbf{\hat{n}}) \quad \text{ for } s=\pm 2.
\end{equation}
Their power spectra are computed as
\begin{equation}
    P_{ss'}(\mathbf{k})=\langle \gamma_{s}(\mathbf{k}) \gamma_{s'}^*(\mathbf{k'}) \rangle ' = \mathbf{\bar{M}}_{ij}^{(s)*}(\mathbf{\hat{n}})\mathbf{\bar{M}}_{kl}^{(s')}(\mathbf{\hat{n}})P_{ijkl}(\mathbf{k});
\end{equation}
where $s,s' =0, \pm 2$. After computing contractions of the type \begin{equation}\label{contr}
    \mathbf{\bar{M}}_{ij}^{(s)*}(\mathbf{\hat{n}})\mathbf{\bar{M}}_{kl}^{(s')}(\mathbf{\hat{n}})\mathbf{Y}_{(ij}^{(m)}(\mathbf{\hat{k}})\mathbf{Y}^{(m)*}_{kl)}(\mathbf{\hat{k}});
\end{equation}
which are only a function of the \textit{angle} $\mu = \mathbf{\hat{n}} \cdot \mathbf{\hat{k}} $ between the wavevector and the line of sight, we obtain expressions for $P_{ss'}$ which are functions of $k$ and $\mu$. Specifically, denoting $s=0, \pm 2$ by resp. $\delta, \pm$ it can be shown that \cite{Kurita_2022}
\begin{equation}\label{pppm}
\begin{aligned}
    P_{\delta +}(k,\mu) = P_{\delta -}(k,\mu) &= \frac{1}{2}\sqrt{\frac{3}{2}}(1-\mu^2)P_{02}^{(0)}(k); \\
    P_{++}(k,\mu) = P_{--}(k,\mu) &= \frac{3}{8}(1-\mu^2)^2P_{22}^{(0)}(k) + \frac{1}{2}(1-\mu^4)P_{22}^{(1)}(k) + \frac{1}{8}(\mu^4+6\mu^2 + 1)P_{22}^{(2)}(k); \\
    P_{+-}(k,\mu) = P_{-+}(k,\mu) &= \frac{1}{8}(1-\mu^2)^2(3P_{22}^{(0)}(k) -4P_{22}^{(1)}(k) +P_{22}^{(2)}(k) ).
\end{aligned}
\end{equation}
In the literature, it is more common to use the `E/B'-basis. The defining relations are 
\begin{equation}
    \gamma_E(\mathbf{k}) := \frac{1}{2}(\gamma_{+2}(\mathbf{k})+\gamma_{-2}(\mathbf{k})); \quad \gamma_B(\mathbf{k}) := \frac{1}{2\mi}(\gamma_{+2}(\mathbf{k})-\gamma_{-2}(\mathbf{k}));
\end{equation}
while the relation to the coordinate-dependent $\gamma_1,\gamma_2$-basis from Eq. \eqref{12} is given by
\begin{equation}
     \gamma_1(\mathbf{k}) := \frac{1}{2}(\gamma_{+2}(\mathbf{k})e^{2\mi \phi_k}+\gamma_{-2}(\mathbf{k})e^{-2\mi \phi_k}); \quad \gamma_2(\mathbf{k}) := \frac{\mi}{2}(\gamma_{+2}(\mathbf{k})e^{2\mi \phi_k}-\gamma_{-2}(\mathbf{k})e^{-2\mi \phi_k}).
\end{equation}
where $\phi_k$ is the azimuthal $\mathbf{k}$-angle, i.e. $\tan{\phi_k} = k_2/k_1$. The spectra for the E- and B-modes are then
\begin{equation}
\begin{aligned}
    P_{EE}(k,\mu) &= \frac{1}{2}(P_{++}(k,\mu)+P_{+-}(k,\mu)); \\
    P_{BB}(k,\mu) &= \frac{1}{2}(P_{++}(k,\mu)-P_{+-}(k,\mu));\\
    P_{\delta E}(k,\mu) &= P_{\delta +}(k,\mu); \\
    P_{\delta B}(k,\mu) &=0.
\end{aligned}
\end{equation}
Thus, the full angular dependence of the $\delta E, EE, BB$-spectra reads\footnote{Note that there is a difference with reference \cite{Kurita_2022} here, as their higher helicity power spectra $P_{22}^{(m)}$ with $m=1,2$ differ from ours by a factor of 2. This is a matter of definition.}
\begin{equation}\label{eebb}
\begin{aligned}
    P_{\delta E}(k,\mu) &= \frac{1}{2}\sqrt{\frac{3}{2}}(1-\mu^2)P_{02}^{(0)}(k); \\
    P_{EE}(k,\mu)&=\frac{3}{8}(1-\mu^2)^2 P_{22}^{(0)}(k) + \frac{1}{2}\mu^2(1-\mu^2) P_{22}^{(1)}(k) + \frac{1}{8}(1+\mu^2)^2 P_{22}^{(2)}(k); \\
    P_{BB}(k,\mu)&= \frac{1}{2}(1-\mu^2) P_{22}^{(1)}(k)+\frac{1}{2}\mu^2P_{22}^{(2)}(k).
\end{aligned}
\end{equation}
When the expressions for the helicity power spectra from Section \ref{iaspec} are inserted, we obtain the theoretical predictions of the EFT of IA. 

Since we only consider the unbiased dark matter density field as scalar tracer, we can simply set all scalar bias parameters equal to zero except for $b_1^s=1$ and $b_R^s$ which is still free. We then decide to fix $b_R^s$ by fitting it to the measured power spectrum $P_{\delta}$ on scales $k<0.2$ $h$/Mpc, yielding $b_R^s = -2.272$. This implies that the spectra in Eq. \eqref{eebb} now depend on $8$ free parameters rather than $14$. We emphasize that fixing $b_R^s$ in this way amounts to a conservative choice with regards to determining the range of validity of the EFT of IA. Fixing it can really only increase $\chi_{\text{red}}^2$ (defined below), because the number of degrees of freedom barely changes if we reduce the number of free parameters by one. 
 
\section{Simulation Data and Statistical Methods}\label{sims}

\subsection{Simulation Data}

The data we use to test the EFT of IA consists of a suite of 20 simulations from the \texttt{DarkQuest} Simulation Project\footnote{\url{https://darkquestcosmology.github.io/}} \cite{Nishimichi19} of size $1\mbox{ (Gpc}/h)^3$ of $2048^3$ dark matter particles, each of which have a mass of $1.02 \times 10^{10}\,h^{-1} M_{\odot}$. We will only consider halos in the mass bin $M_{\rm h} = [10^{12}, 10^{12.5}] \,h^{-1} M_{\odot}$, i.e. each halo contains $\mathcal{O}(100)$ particles. This sample has the highest signal-to-noise intrinsic alignment multipoles in the simulations. We restrict ourselves to redshift $z=0$ here. Details on how the halo shape inertia tensor is constructed and how the ellipticity field is defined can be found in \cite{Kurita_2020}.

The angular dependence of the projected shape power spectra $P_{XY}(k,\mu)$ can be captured in their multipole moments, which are computed from the simulation as 

\begin{equation}\label{est}
    \hat{P}_{XY}^{\ell}(k_i) := \frac{2\ell+1}{N_i}\sum_{\mathbf{k} \in \textrm{shell}~i}\mathcal{L}_{\ell}(\mu) X(\mathbf{k})Y(-\mathbf{k}); \quad \text{where } k_i=\frac{1}{N_i}\sum_{\mathbf{k} \in \textrm{shell}~i}|\mathbf{k}|.
\end{equation}

Here $X,Y = \delta, E, B$ while $\mathcal{L}_{\ell}(\mu)$ is a Legendre polynomial and $N_i$ is the total number of modes in the shell.  The Fourier-transformed fields 
$X(\mathbf{k})$ are used to compute the power spectrum multipoles in $40$ equally linearly spaced bins 
over $0<k<1\,h$/Mpc.
The line of sight direction is taken to be the $z$-axis, so that $\mu = \mathbf{k}_3 / k$. Here, for simplicity, we only consider multipole moments $\ell=0,2$, i.e. the monopole and quadrupole moments\footnote{In principle, it is possible to also include the hexadecapole $P_{EE}^{(4)}$, but it turns out that its signal-to-noise ratio is low compared to the lower multipoles.}. 

The estimator in Eq. \eqref{est} satisfies 
\begin{equation}\label{estavg}
   \langle\hat{P}_{XY}^{\ell}(k_i)\rangle = \frac{2\ell+1}{N_i}\sum_{\mathbf{k} \in \textrm{shell } i}\mathcal{L}_{\ell}(\mu)P_{XY}(k,\mu), 
\end{equation}
where the brackets denote ensemble averaging. Hence, it can be compared to theory by computing the r.h.s. using Eq. \eqref{eebb}. This requires computing the theory prediction for every mode $\mathbf{k}$ in the Fourier space lattice (see also \cite{Nishimichi_2020}), which is very slow if the loop integrals are to be computed numerically. To overcome this, we first computed the integrals numerically on a coarser grid and then interpolated the result. We used the \texttt{Cuba} library \cite{Hahn_2005} for numerical integration in \texttt{Mathematica} v12.0 \cite{math}. By integrating Eq. \eqref{eebb} over $\mu$, we observe that the stochastic component $P_{\epsilon}^g$, which only occurs in the monopoles of the E- and B-mode autospectra, must be equal for both. This prediction is not based on the EFT specifically, but rather on symmetry arguments. We test this assumption by leaving $c^g$ free.

\subsection{Covariance Modelling}

To model the covariance of the multipole power spectra, we will make the assumption that the underlying fields are Gaussian. Then one can show that \cite{Guzik_2010,Nishimichi_2020} 
\begin{equation}\label{cov}
\begin{aligned}
    \textrm{Cov}\bigg(\hat{P}_{XY}^{\ell}(k_i),\hat{P}_{X'Y'}^{\ell'}(k_j)\bigg) &= \frac{(2\ell+1)(2\ell'+1)}{N_i^2}\delta_{ij} \\ &\times \sum_{\mathbf{k} \in \textrm{shell } i}\mathcal{L}_{\ell}(\mu)\mathcal{L}_{\ell'}(\mu)\bigg( P_{XX'}(\mathbf{k})P_{YY'}(\mathbf{k})+P_{X'Y}(\mathbf{k})P_{XY'}(\mathbf{k}) \bigg)
\end{aligned}
\end{equation}
It turned out that using the measured power spectra for the covariance estimate was too noisy given the relatively low number of realizations. As an alternative, we will assume the nonlinear alignment model for the power spectra $P_{XY}$ that enter the covariance in Eq. \eqref{cov}. More precisely, in computing the covariance we make the ansatz
\begin{equation}\label{nla}
    \begin{aligned}
        P_{EE}(k,\mu) &= \frac{1}{4}(b_1^g)^2(1-\mu^2)^2P_\delta(k) + c^g; \\
        P_{\delta E}(k,\mu) &= \frac{1}{2}b_1^g(1-\mu^2)P_\delta(k); \\
        P_{BB}(k,\mu) &= c^g.
    \end{aligned}
\end{equation}
Here $P_{\delta}$ is the theoretical, \textit{fully nonlinear} matter power spectrum. Thus, the covariance now depends on the two unknown parameters $b_1^g$ and $c^g$. In the linear regime, i.e. $k<0.05$ $h$/Mpc, the linear alignment model as well as the Gaussianity assumption are certainly expected to hold. For this reason, we can simultaneously determine fiducial values for $b_1^g$ and $c^g$ by starting from some initial values $b_1^{g*}, c^{g*}$ and then computing (i) the covariance matrix, (ii) the best fit values $b_1^g{}', c^g{}'$ determined by using this covariance when fitting $P_{\delta E, EE, BB}^{(0,2)}$ in the regime $k<0.05\,h$/Mpc and finally (iii) recomputing the covariance matrix using $b_1^g{}', c^g{}'$ and repeating the above steps until the process stabilizes. In practice, this happens already after a few ($<10$) steps. The covariance thus obtained is then used when fitting the EFT of IA to the simulation data in the quasi-linear regime. We denote the ground truth value for $b_1^g$ by $b_1^f$ and similarly the ground truth for $c^g$ by $c^f$. The $1\sigma$ errors $\sigma_{b^f}, \sigma_{c^f}$ on $b_1^f$, $c^f$ can be obtained by using standard sampling methods. We find 
\begin{equation}\label{fid}
\begin{aligned}
    b_1^f &= -0.0776; & \sigma_{b^f} &= 0.0021;\\
    c^f &=9.66 \text{ (Mpc}/h)^3; & \sigma_{c^f} &= 0.25  \text{ (Mpc}/h)^3.
\end{aligned}
\end{equation}
It is also possible to determine the ground truth $b_1^f$ for the bias parameter $b_1^g$ on a realization-by-realization basis, i.e. by considering the ratio of measured power spectra $P_{\delta E,r} / P_{\delta,r}$ in the linear regime ($k<0.05$ $h$/Mpc) for every realization $r$ and fitting a constant value $(b_1^g) _r$ to it. Then, since all realizations are independent, the fiducial value for $b_1^g$ is taken to be the sample mean and the error $\sigma_f$ is the sample error over all realizations. This method was applied in \cite{Kurita_2020} and we have checked that our method is consistent with it. 

In general, the covariance matrix receives contributions from (i) the Gaussian part discussed above; (ii) a `connected non-Gaussian' part; and (iii) the super-sample covariance \cite{Takada_2013,Barreira18,Kurita_2020}, which however is absent in simulation data, as there are no modes with wavelength larger than the fundamental of the box. One could improve the accuracy of the covariance matrix by incorporating the connected non-Gaussian part perturbatively \cite{Wadekar_2020}, or by using Eq.  \eqref{eebb} rather than Eq. \eqref{nla}. 

\subsection{Goodness of Fit and Figure of Merit} 
As a measure of goodness of fit, we employ the chi-squared statistic which is given by the Gaussian log-likelihood:
\begin{equation}\label{chisqstat}
\begin{aligned}
    \chi^2(k_{\text{max}}) &= \frac{1}{N_R} \sum_{r=1}^{N_R}\sum_{\substack{XY,X'Y', \\i,j,\ell,\ell'}} \bigg([\hat{P}_{XY}^{\ell}(k_i)]_r- P_{XY}^{\ell}(k_i)_{\text{th}}\bigg)\\
    &\times \bigg[\textrm{Cov}\bigg(\hat{P}_{XY}^{\ell}(k_i),\hat{P}_{X'Y'}^{\ell'}(k_j)\bigg) \bigg]^{-1} \bigg([\hat{P}_{X'Y'}^{\ell'}(k_j)]_r- P_{X'Y'}^{\ell'}(k_j)_{\text{th}} \bigg).
\end{aligned}
\end{equation}
Equivalently, we define the likelihood for the collection of independent boxes to be the geometric mean of the individual likelihoods. As such, the presented analysis can be interpreted as computing the $k_{\text{max}}$ up to which the models are valid in a volume equal to that of a single box. Here $P_{XY}^{\ell}(k_i)_{\text{th}} = \langle\hat{P}_{XY}^{\ell}(k_i)\rangle$ is the theory estimate described above and $N_R = 20$ is the number of realizations. The quantity $[\hat{P}_{XY}^{\ell}(k_i)]_r$ is the measured power spectrum multipole in the $r$-th realization and the covariance is given by Eq. \eqref{cov}. The sum over the bins $i,j$ runs over all bins which satisfy $k_i<k_{\text{max}}$ and $XY, X'Y'$ can be $\delta E, EE, BB$ while $\ell, \ell' = 0,2$. Note that we use the same covariance for every realization. 

We determine the best-fit values of the $8$ free bias parameters for a given $(k_{\text{max}})$ by minimizing $\chi^2(k_{\text{max}})$. It can be shown that the expression
\begin{equation}\label{barchi}
\begin{aligned}
    \overline{\chi}^2(k_{\text{max}}) &= \sum_{\substack{XY,X'Y', \\i,j,\ell,\ell'}} \bigg(\overline{\hat{P}_{XY}^{\ell}(k_i)}- P_{XY}^{\ell}(k_i)_{\text{th}}\bigg) \\
    &\times \bigg[\textrm{Cov}\bigg(\hat{P}_{XY}^{\ell}(k_i),\hat{P}_{X'Y'}^{\ell'}(k_j)\bigg) \bigg]^{-1} \bigg(\overline{\hat{P}_{X'Y'}^{\ell'}(k_j)}- P_{X'Y'}^{\ell'}(k_j)_{\text{th}} \bigg)
\end{aligned}
\end{equation}
differs only from Eq. \eqref{chisqstat} by a constant \cite{Eggemeier_2020}, where
\begin{equation}
    \overline{\hat{P}_{XY}^{\ell}(k_i)}=\frac{1}{N_R}\sum_{r=1}^{N_R}[\hat{P}_{XY}^{\ell}(k_i)]_r
\end{equation}
is the sample average of the measured multipoles. Hence minimizing Eq. \eqref{barchi} is equivalent to minimizing Eq. \eqref{chisqstat}, but Eq. \eqref{barchi} is more convenient as it does not involve a sum over all realizations. 

\begin{table}[]
\resizebox{\textwidth}{!}{%
    \centering
    \begin{tabular}{|c|c|c|c|c|c|c|c|}
    \hline
        Parameter & NLA & TATT & TATT+VS & 6p-EFT (LP) & 6p-EFT &EFT(LP) & EFT\\
        \hline
        \hline
        $b_1^g$ & [-0.3,0] & [-0.3,0] & [-0.3,0] & [-0.3,0] & [-0.3,0] & [-0.3,0] & [-0.3,0] \\
        $b_{2,1}^g$ & n.a. &[-3,3] & [-3,3] & [-3,3] & [-3,3] & [-3,3] & [-3,3] \\
        $b_{2,2}^g$ & n.a. & [-3,3] & [-3,3] & [0,3] & [-3,3] & [0,3] & [-3,3]\\
        $b_{2,3}^g$ & n.a. & n.a. & [-3,3] & [-3,0] & [-3,3] & [-3,0] & [-3,3]\\
        $b_{3,1}^g$ & n.a. & n.a. & n.a. & n.a. & n.a. & [-3,3] & [-3,3]\\
        $b_{3,2}^g$ & n.a. & n.a. & n.a. & n.a. & n.a. & [-3,3] & [-3,3]\\
        $b_{R}^g$ & n.a. & n.a. & n.a. & [-3,3] & [-3,3] & [-3,3] & [-3,3]\\
        $c^g$ [(Mpc/$h$)$^3$] & [7,13] & [7,13] & [7,13] & [7,13] & [7,13]& [7,13] & [7,13]\\
        \hline
        Total & 2 & 4 & 5 & 6 & 6 & 8 & 8\\
        \hline
    \end{tabular}}
    \caption{Priors for free parameters in each of the intrinsic alignment models. All prior distributions are uniform on the given interval. Note that all models, the stochastic parameter is still free. Bottom row indicates the total number of free parameters. Here `LP' stands for Lagrangian prior as described in Section \ref{models} and `n.a.' means that the parameter does not occur in the model, i.e. in the TATT model we have opted to use $b_{2,2}^g$.}
    \label{tab:tab1}
\end{table}
For a given $k_{\text{max}}$, the number of degrees of freedom for a given model equals 
\begin{equation}
\begin{aligned}
     N_\text{dof}(k_{\text{max}}) &= (\# \text{ multipoles})\times N_R \times (\# \text{ bins }k_i < k_{\text{max}}) - ( \# \text{ free parameters}) \\
     &= 120 \times (\# \text{ bins }k_i < k_{\text{max}}) - ( \# \text{ free parameters}).
\end{aligned}
\end{equation}
As a goodness of fit statistic, we then compute $\chi_{\text{red}}^2(k_{\text{max}})=\chi^2 N_R/ ( N_\text{dof})$ as a function of $k_{\text{max}}$\footnote{We expect $\chi^2 N_R$ to follow a chi-squared distribution with $N_{\text{dof}}$ degrees of freedom, which is approximately Gaussian when $ N_\text{dof} \gg 1$, which indeed applies here.}. 
We will do this for the NLA model, the EFT of IA and all the models described in Section \ref{models}. 

An additional requirement for the a model to be a good fit is to correctly reproduce the ground truth value $b_1^f$ of $b_1^g$, the linear alignment parameter\footnote{The same holds for the stochastic noise parameter $c^g$, but it turns out that including this in the requirement does not make a significant difference as for determining $k_{\text{max}}$.}. More specifically, we can define the figure of bias (FoB) as 
\begin{equation}
    \text{FoB}(k_{\text{max}}) = \frac{|b_1^g(k_{\text{max}}) - b_1^f|}{\sqrt{\sigma(k_{\text{max}})^2 + \sigma_{b^f}^2}};
\end{equation}
where $\sigma(k_{\text{max}})$ is the $1\sigma$ posterior error on $b_1^g(k_{\text{max}})$ and $b_1^f, \sigma_{b^f}$ are the ground truths defined above. For the EFT to be successful in describing the intrinsic alignment power spectra, both the FoB and $\chi_{\text{red}}^2$ should not differ too much from unity. We will employ the criterion used in \cite{Eggemeier_2020}, namely 
\begin{equation}\label{crit}
    \Delta(k_{\text{max}}) = \text{FoB}(k_{\text{max}}) + \frac{\chi_{\text{red}}^2 -1 }{2\sigma_{\text{red}}} < \Delta_{c} = 1.5;
\end{equation}
where $\sigma_{\text{red}} = \sqrt{2/N_\text{dof}}$ equals one standard deviation of the distribution of $\chi^2_{\text{red}}$. This determines $k_{\text{max}}$ as the largest wavenumber for which Eq. \eqref{crit} holds, which we then determine for all the aforementioned models.

Lastly, we also determine for all models the \textit{figure of merit} (FoM) defined by
\begin{equation}
    \text{FoM}(k_{\text{max}}) = \frac{|b_1^g(k_{\text{max}})|}{\sigma(k_{\text{max}})};
\end{equation}
which quantifies how well the model is able to place constraints on the linear bias parameter. 
After determining the best-fit EFT bias parameters, we also access their posterior probability distribution by using the Python Dynamic Nested Sampling library \texttt{dynesty} \cite{Speagle_2020,2004AIPC..735..395S,2009MNRAS.398.1601F,Skilling_06,Higson_2018,koposov_22}\footnote{In fact, we can use Eq. \eqref{barchi} for computing the log-likelihood $-2 \log \mathcal{L} = \chi^2$ rather than Eq. \eqref{chisqstat}.}. We adopt wide uniform priors as listed in Table \ref{tab:tab1}. We use 500 live points and a random walk scheme for determining the proposals at each iteration. 

For determining the $k_{\text{max}}$ at which the model breaks down it is sufficient to continue sampling up to $8\times 10^4$ effective samples\footnote{With the exception of the EFT models, where we continue sampling up to $1.6 \times 10^5$ effective samples.}. This also suffices to determine the error bars on the other parameters in the cases where the Lagrangian prior is imposed. However, due to the quadratic dependence on the second order bias parameters of any of these models (except NLA), the posterior distribution for these bias parameters displays strongly bimodal features if no priors are imposed. We were able to produce a reliable corner plot in the case of the full EFT of IA with Lagrangian prior imposed, where we found that it is necessary to continue sampling up to $1 \times 10^6$ effective samples\footnote{We run several independent instances of the sampler and compare the outputs to ensure that the posterior has stabilized.}. 
\section{Results}\label{results}
\begin{figure}[h!]
\centering
\includegraphics[width=\linewidth]{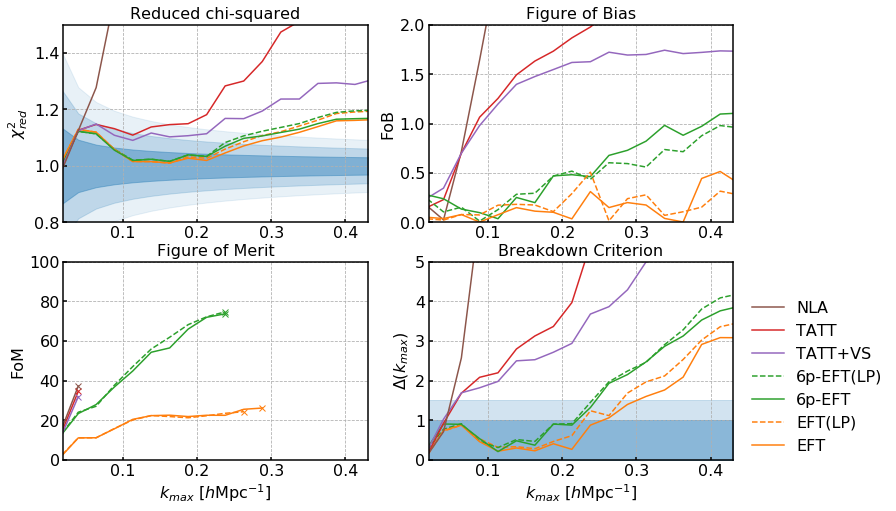}
\caption{Comparison of several different alignment models in the quasi-linear regime. \textit{Top Left:} $\chi^2_{\text{red}}$ as a function of $k_{\text{max}}$ for the EFT of IA. Dark blue indicates the $1\sigma$ bound on $\chi^2_{\text{red}}$, i.e. $1 \pm \sqrt{2/ N_\text{dof}}$. The lighter blue regions indicate the $2\sigma$ and $3\sigma$ bounds. Note that when the curve exits the $3\sigma$ bound, \eqref{crit} is always violated. \textit{Top Right}: Figure of bias for all alignment models considered. \textit{Bottom Left}: Figure of merit. We stop plotting the figure of merit beyond $k_{\rm max}$, which is indicated by the cross for each model. \textit{Bottom Right:} $\Delta(k_{\text{max}})$ for the same range of values of $k_{\text{max}}$ as the previous plot. Dark blue indicates $\Delta < 1$ while light blue indicates $\Delta < \Delta_{c} = 1.5$. }. 
\label{fig:res1}
\end{figure}

We first show the $\chi_{\text{red}}^2, \Delta$, the figure of merit and figure of bias as a function of $k_{\text{max}}$ for the models described above in Figure \ref{fig:res1}. 
We infer that the EFT of IA fulfils our criterion from Eq. \eqref{crit} up to $k_{\text{max}}=0.30\,h$/Mpc (or $k_{\text{max}}=0.28\,h$/Mpc with Lagrangian prior (LP, cf. Section \ref{models}). The 6-parameter model is valid up to $k_{\text{max}}=0.25\,h$/Mpc regardless of the prior. However, the analysis for the NLA, TATT and TATT+VS models shows that Eq. \eqref{crit} is only valid up to $k_{\text{max}}=0.05\,h$/Mpc (for this reason, we do not consider the LP for these models). Imposing the LP on either the 6-parameter model or the full EFT does not heavily impact their performance on the data, but has the additional advantage that it becomes simpler to summarize the posterior distributions of $b_{2,2}^g$ and $b_{2,3}^g$. For this reason, we restrict ourselves to the LP from now on for these two models. Note that when comparing the several different alignment models, it is important to include the NLA model in order to ensure that one is not just inadvertently modeling nonlinearities in the matter power spectrum using the additional free bias parameters. Indeed, if the goodness of fit using the NLA model were comparable to that of the more general bias expansions, then the additional bias parameters are simply capturing additional corrections to the matter power spectrum beyond one-loop rather than dependencies on any of the higher order operators.

The breakdown of the NLA, TATT and TATT+VS models seems to be due to a combination of both higher figure of bias and higher $\chi_{\text{red}}^2$. As expected, all models generally show an increase in figure of merit with $k_{\text{max}}$. The 6-parameter model is able to obtain the tightest constraint on $b_1^g$ within its range of validity. 

\begin{figure}[h!]
    \centering
    \includegraphics[width=\textwidth]{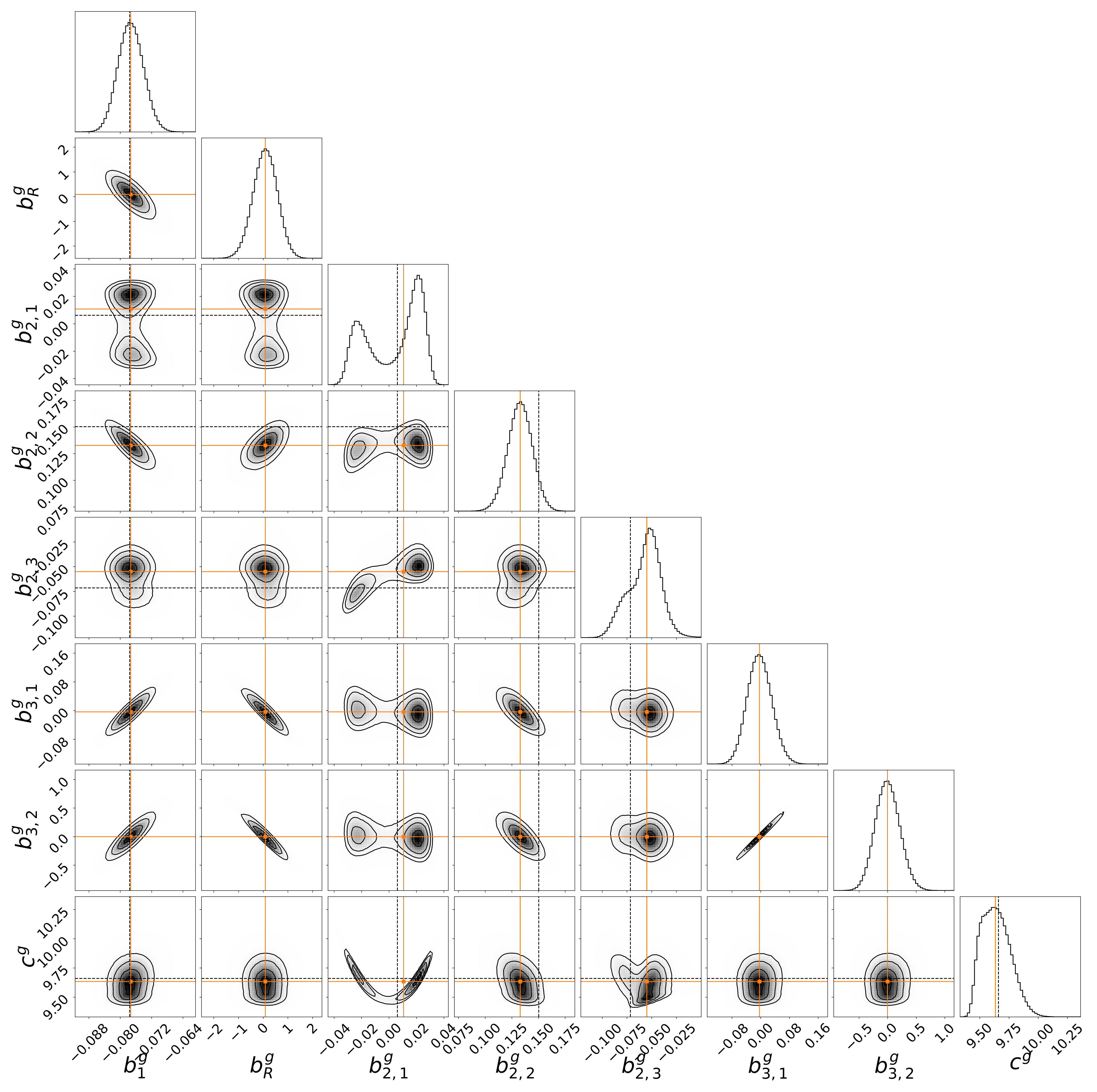}
    \caption{Posterior distributions for all shape bias parameters for the EFT of IA with Lagrangian prior imposed on $b_{2,2}^g$ and $b_{2,3}^g$ at $k_{\text{max}} = 0.28 \,h$/Mpc. Orange lines indicate posterior mean, while the black dashed lines indicate `fiducial' values, i.e. the true values for $b_1^g$ and $c^g$ from Eq. \eqref{fid} and the Lagrangian predictions for $b_{2,1}^g, b_{2,2}^g, b_{2,3}^g$ from Eq. \eqref{lagr}.}
    \label{fig:mcmc}
\end{figure}

\begin{figure}
\centering
\includegraphics[width=\linewidth]{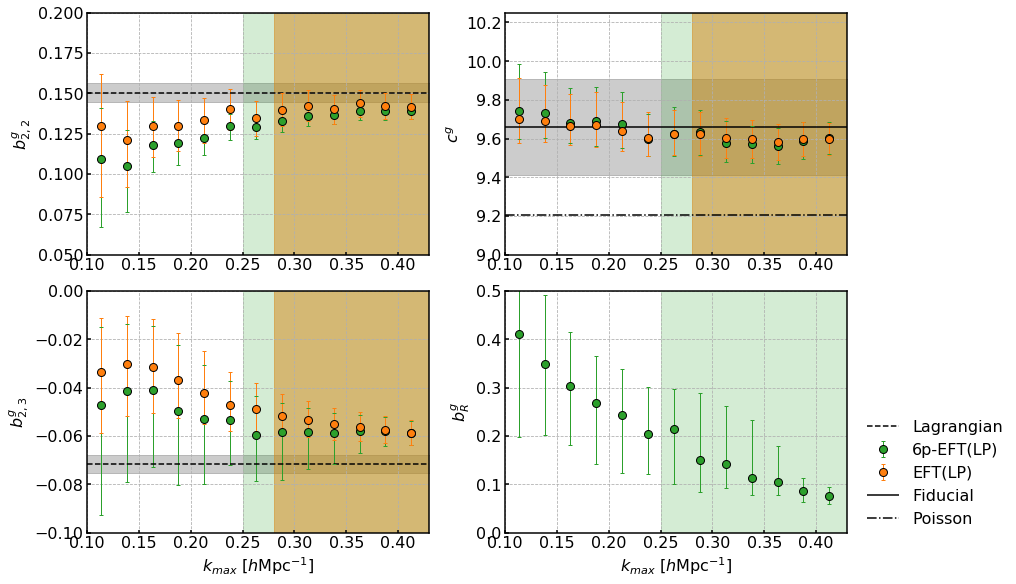}
\caption{Marginalized posterior mean and standard deviation of bias parameters $b_{2,2}^g, b_{2,3}^g, c^g$ and $b_R^g$ for the 6-parameter model and full EFT (except for $b_R^g$) with the Lagrangian prior imposed. Green and orange shaded bands mark the regime where the 6-parameter model resp. the full EFT breaks down. Grey horizontal shaded bands indicate uncertainty on the dashed parameters. Dot-dashed line in the bottom left plot shows the Poisson noise $\sigma_{\gamma}^2/\bar{n}_h$ as measured from randomizing the orientation of the shapes \cite{Kurita_2020}.  Error bars indicate the 16th and 84th percentiles of the posterior distribution. Other parameters are consistent with zero and hence are not shown.}
\label{fig:res2}
\end{figure}

\begin{figure}
    \centering
    \includegraphics[width=\textwidth]{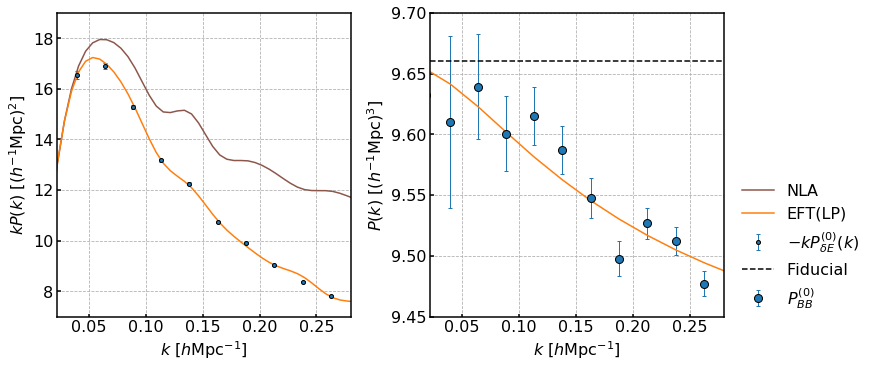}
    \caption{Best fit to the monopoles $P_{\delta E}^{(0)}$ (left) and $P_{BB}^{(0)}$ (right) with the EFT of IA (orange curve) with Lagrangian prior imposed for $k_{\text{max}}=0.28\,h$/Mpc. The data points correspond to the mean value over all $N_R=20$ realizations and the errors are rescaled to the mean, i.e. divided by $\sqrt{N_R}$. We also show the fiducial shot noise value from Eq. \eqref{fid} on the right and the best-fit NLA curve on the left fitted on linear scales $k<0.05\,h$/Mpc. Note that the error bars in the left plot are there, but barely visible.}
    \label{fig:bb}
\end{figure}

Next, we compute the posterior distributions of all EFT bias parameters including all $k$-bins with $k\leq k_{\text{max}}=0.28\,h$/Mpc. The result is shown in Figure \ref{fig:mcmc}. 
A number of things can be seen from this plot: 
\begin{itemize}
    \item There is a degeneracy between $b_R^g, b_{3,1}^g, b_{3,2}^g$, as alluded to in Section \ref{models}. While none of these bias parameters are detected at any significance, this does not mean that one can put them equal to zero without degrading the quality of the fit. Indeed, the \textit{joint} posterior distribution of the triple $(b^g_{3,1},b^g_{3,2},b_R^g)$ is localized away from the origin. This is also exemplified by the bottom right plot of Figure \ref{fig:res2}, where the single third-order bias parameter for the 6-parameter model is detected at mild significance ($\sim 2.5\sigma$). 
    \item Moreover, $b_1^g$ is also correlated with the third-order parameters. Setting these parameters to zero, as is done in all cases except the EFT and the 6-parameter model, drives $b_1^g$ away from its fiducal value and explains the increased figure of bias and $\chi_{\text{red}}^2$ for NLA, TATT and TATT+VS. Put differently, setting $b_R^g = 0$ leads to artificially tight constraints on $b_1^g$ around the wrong value, as is seen from the figure of merit and figure of bias plots. Indeed, the two-dimensional joint posterior for $(b_1^g, b_R^g)$ in the 6-parameter model reveals a significant anti-correlation between the two. 
    \item The relation $b_{2,2}^g=b_{2,3}^g$ is strongly disfavoured by the EFT at this scale, which is in accordance with the higher value of $\chi^2_{\text{red}}$ for the TATT model. 
    \item We see that the stochastic amplitude $c^g$ is correlated with $(b_{2,1}^g)^2$. This can be understood by considering $P_{BB}^{(0)}$ (cf. Figure \ref{fig:bb}), which shows a clear downward trend and depends both on $c^g$ and on $(b_{2,1}^g)^2$ via the $I_{55}$ and $I_{66}$ terms in Eq. \eqref{eftpowspec}. These two loop integrals show more or less the same downward trend, as seen from Figure \ref{fig:iplot} (they are negative). Thus, when $c^g$ is low, the slope of the best-fit curve and thus the value of $(b_{2,1}^g)^2$ also tends to be low, and vice versa. 
\end{itemize} 
The posterior distributions of $b_{2,2}^g, b_{2,3}^g, c^g$ and the single third order bias parameter $b_R^g$ for the 6-parameter model as a function of $k_{\text{max}}$ are shown in Figure \ref{fig:res2}. 

As expected, we do not see any significant scale dependence (i.e. `running' with $k_{\text{max}}$) for any of the bias parameters. The resulting parameter constraints at the largest admissible value of $k_{\text{max}}$ are given in Table \ref{tab:tab2}. 

From these results, it appears that there is at least some tension between the posterior distribution of the EFT model parameters and the Lagrangian prior values, but not to the extent that we are forced to conclude that the choice of \textit{sign} of $b_{2,2}^g, b_{2,3}^g$ is \textit{a posteriori} unjustified.
\begin{table}[]
    \centering
    \begin{tabular}{|c|c|c|}
    \hline
        Parameter & 6p-EFT (LP) & EFT (LP)\\
        \hline\hline
        $b_{2,2}^g$ & $0.129 ^{+0.008}_{-0.008}$ & $0.135 ^{+0.011}_{-0.011}$\\
        $b_{2,3}^g$ & $-0.054 _{-0.018} ^{+0.016}$  & $-0.049^{+0.011}_{-0.009}$\\
        $c^g\ [\text{(Mpc/h)}^3]$ & $9.60 ^{+0.13}_{-0.09}$ & $9.62 ^{+0.13}_{-0.11}$\\
        $b_R^g$ & $0.20 ^{+0.10} _{-0.08}$ & n.a.  \\
        \hline
    \end{tabular}
    \caption{Posterior values for $c^g, b^g_{2,2}, b^g_{2,3}$ for the 6-parameter model and full EFT evaluated at their largest admissible $k_{\text{max}}$.}
    \label{tab:tab2}
\end{table}
It is also interesting to note that the Lagrangian prior implies $b_{KK} = b_{2,1}^g-b_{2,3}^g = -b_1^g > 0$. Then, $b_1^g$ and $b_{KK}$ have opposite signs, so if linear response is radial, then tidal torquing is tangential. However, the full EFT posterior for $b_{2,3}^g$ (without the Lagrangian prior imposed) is roughly symmetric around the origin, thus making it possible for $b_{2,3}^g$ to be positive so that $b_{KK} < 0$. It thus remains unclear for this sample of halos what sign of tidal torquing is preferred. It is also interesting to note that generally shape alignments of halos might not have the same sign as their angular momentum alignments, and that this varies with mass and redshift \cite[e.g.,][]{Codis,Lopez,Moon}. Therefore, a more in-depth study of how $b_{KK}$ evolves with mass and redshift, and an extension of the EFT to model the angular momentum of halos would be of interest. 

Lastly, we show the best fit curves from the EFT of IA (with the Lagrangian prior imposed) up to $k_{\text{max}}=0.28\,h$/Mpc for $P_{BB}^{(0)}$ and $P_{\delta E}^{(0)}$ in Figure \ref{fig:bb}. From this figure, it is clear that for $k \gtrsim 0.15\,h$/Mpc, the BB monopole is different from just the shot noise signal and the EFT of IA is able to consistently describe it. Further, the $\delta E$ monopole, which is measured with very high signal to noise, is described very poorly by the NLA model for $k>0.05\,h$/Mpc, in contrast to the EFT of IA. 

\section{Discussion} 
We now discuss some caveats and possible extensions of our analysis.

In this work, we did not include the effect of the damping of the Baryon Acoustic Oscillation (BAO) feature in the model predictions. This can be calculated by means of resummation in perturbation theory \cite{Vlah20,Baldauf_2015,Vlah:2015,Blas_2016}, and including it could improve the quality of the fits further, given that the feature was identified in the alignment spectra in \cite{Kurita_2020}.  

One might also want to consider the inclusion of higher-derivative effects on intrinsic shapes by including the corresponding operators in the field expansion \cite{Vlah20}. These might be important for more massive halo samples. This applies additionally to stochastic effects beyond leading order. For example, the authors of \cite{Eggemeier_2020} consider subleading stochastic contributions when modeling the power spectrum for galaxy clustering. Given the recent interest in the bispectrum as a means of adding constraining power for cosmological inference \cite{Eggemeier_2021,moretti22,Damico22}, a similar effort could be made for bispectra involving shape fields using the tree-level expressions from \cite{Vlah20} (see also \cite{Pyne_2022}). The addition of the bispectrum is expected to break degeneracies in the second order bias parameters, since the bispectrum contains exactly one second-order field at tree-level. It does, however, also introduce more stochastic parameters. 

Finally, in order to justify applying the EFT of IA for cosmological inference, one should also (i) verify that the increase in $k_{\text{max}}$ for EFT of IA translates into tighter cosmological constraints compared to simpler models and (ii) model redshift space distortions (RSD) consistently for shapes for applications to galaxy surveys rather than simulations. We leave these issues for future work.  

For $k \lesssim 0.2\,h$/Mpc, one can see from the analysis of \cite{Kurita_2020} that the Gaussian part of the covariance matrix dominates in the signal-to-noise calculation. The contributions from the connected non-Gaussian part and the super-sample covariance bring the signal-to-noise down, which suggests that our assumption underestimates the measurement errors and thus yields a conservative value for $k_{\text{max}}$. However, the NLA model from Eq. \eqref{nla} \textit{overestimates} the $EE$ and $\delta E$ spectra (see for instance the left panel of Figure \ref{fig:bb}) and thus the error bars. Since $\chi^2_{\text{red}}$ is quite sensitive to the covariance, there is still considerable uncertainty regarding the range of validity of the higher order models. If one were to employ a more accurate and robust model for the covariance, the width of the posteriors for the bias parameters would also change at a given scale cut $k_{\text{max}}$. This could be achieved through a larger suite of high-resolution simulations than was available for this work. In any case, more future work on validating modelling choices for the shape covariance in the quasi-linear regime $k \gtrsim 0.1 \,h$/Mpc is certainly necessary. 

We end our discussion with some cautionary remarks regarding the interpretation of our results. We stress that this analysis is done at $z=0$ for halos of a specific mass. It is therefore as of yet unclear how these results can be extrapolated to more realistic redshift and mass ranges of upcoming spectroscopic surveys like the Dark Energy Spectroscopic Instrument\footnote{\url{https://www.desi.lbl.gov/}} and to galaxies. Nevertheless, we emphasise that the EFT approach is agnostic about halo mass and redshift: this dependence is accounted for in the free bias parameters of the theory. Similarly, galaxy shapes can be decomposed in the same basis of operators, so that the EFT predictions described here could be applied directly to galaxy shapes as well.

Additionally, it remains to be seen whether the EFT of IA is viable for mitigation purposes in weak lensing surveys (see Sec. VII.B in \cite{Secco22} or Sec. VII in \cite{Amon_2022}). In principle, as shown here, the theory breaks down at quasi-linear scales. However, current weak lensing surveys have less constraining power over the alignment model compared to the simulation we use in this work. With lensing studies typically trying to extract information from nonlinear scales, using the EFT of IA would certainly be an extrapolation. How this would impact cosmological constraints remains to be understood. From a physical perspective, in this regime, a halo model-based approach as presented in \cite{Fortuna21b} might be a better option. We leave testing the EFT of IA in the context of lensing mitigation for future work. 

\section{Conclusions}
We have presented the predictions for 3D shape power spectra in the EFT of IA, correcting some mistakes in equations of previously published work, and performed a comparison to measured statistics for halos in the mass range of $M_{\rm h} = 10^{12}-10^{12.5} \,h^{-1} M_{\odot}$ identified in dark-matter-only simulations. 

We have demonstrated that the theory yields an excellent fit to simulations in the range of scales $k \lesssim 0.3\,h$/Mpc. We have identified two beyond-linear bias parameters that are detected to high significance in the alignments of halos up to this scale. In comparison, under the same goodness-of-fit criterion, the NLA model, which has only two free parameters, only offers similar accuracy up to $k=0.05\,h$/Mpc. The same conclusion applies to the TATT model, either with or without the inclusion of velocity shear. 

In addition, the EFT of IA correctly predicts the $B$-mode power spectrum of intrinsic alignments, which are detected significantly in the simulations.

Our work opens up the possibility to use the EFT of IA in the context of extracting cosmological information from alignment observables, and of mitigating the role of IA as a weak lensing and clustering contaminant. 

\acknowledgments
This publication is part of the project ``A rising tide: Galaxy intrinsic alignments as a new probe of cosmology and galaxy evolution'' (with project number VI.Vidi.203.011) of the Talent programme Vidi which is (partly) financed by the Dutch Research Council (NWO). This work is also part of the Delta ITP consortium, a program of the Netherlands Organisation for Scientific Research (NWO) that is funded by the Dutch Ministry of Education, Culture and Science (OCW).
We thank Marko Simonovic, Alex Eggemeier and Elena Sellentin for useful discussions. We acknowledge the Lorentz Center and the hol-IA workshop\footnote{\url{https://www.lorentzcenter.nl/hol-ia-a-holistic-approach-to-galaxy-intrinsic-alignments.html}} for support while this work was in its final stages.
TK is supported by Research Fellowships of the Japan Society for the Promotion of Science (JSPS) for Young Scientists. 
\vspace{2cm}

\appendix
\section{Perturbation Theory Kernels}\label{kernels}
The perturbation theory kernels for the (22) part of the power spectrum are
\begin{equation}\label{ikernels}
\begin{aligned}
K^I_{11}(\mathbf{u},\mathbf{v}) &= \bigg(\frac{5}{7}+\frac{1}{2}(\mathbf{u}\cdot \mathbf{v})(\frac{1}{u^2}+\frac{1}{v^2})+\frac{2}{7}\frac{(\mathbf{u}\cdot \mathbf{v})^2}{u^2v^2}\bigg)^2; \\
K^I_{12}(\mathbf{u},\mathbf{v}) &= \frac{(\mathbf{u}\cdot \mathbf{v})^2}{u^2 v^2} \cdot \bigg(\frac{5}{7}+\frac{1}{2}(\mathbf{u}\cdot \mathbf{v})(\frac{1}{u^2}+\frac{1}{v^2})+\frac{2}{7}\frac{(\mathbf{u}\cdot \mathbf{v})^2}{u^2v^2}\bigg);\\
K^I_{13}(\mathbf{u},\mathbf{v}) &= \frac{5}{7}+\frac{1}{2}(\mathbf{u}\cdot \mathbf{v})(\frac{1}{u^2}+\frac{1}{v^2})+\frac{2}{7}\frac{(\mathbf{u}\cdot \mathbf{v})^2}{u^2v^2};\\
K^I_{22}(\mathbf{u},\mathbf{v}) &= \frac{(\mathbf{u}\cdot \mathbf{v})^4}{u^4 v^4}; \\
K^I_{23}(\mathbf{u},\mathbf{v}) &= \frac{(\mathbf{u}\cdot \mathbf{v})^2}{u^2 v^2};\\
K^I_{24}(\mathbf{u},\mathbf{v}) &= \frac{(\mathbf{u}\cdot \mathbf{v})^3(2(\mathbf{u}\cdot \mathbf{v})(u^2+v^2)+3u^2v^2+(\mathbf{u}\cdot \mathbf{v})^2)}{\sqrt{6}u^4 v^4 |\mathbf{u+v}|^2};\\
K^I_{33}(\mathbf{u},\mathbf{v}) &= 1; \\
K^I_{34}(\mathbf{u},\mathbf{v}) &= \frac{(\mathbf{u}\cdot \mathbf{v})(2(\mathbf{u}\cdot \mathbf{v})(u^2+v^2)+3u^2v^2+(\mathbf{u}\cdot \mathbf{v})^2)}{\sqrt{6}u^2 v^2 |\mathbf{u+v}|^2};\\
K^I_{44}(\mathbf{u},\mathbf{v}) &= \frac{(\mathbf{u}\cdot \mathbf{v})^2\bigg(2(\mathbf{u}\cdot \mathbf{v})(u^2+v^2)+3u^2v^2+(\mathbf{u}\cdot \mathbf{v})^2\bigg)^2}{6u^4 v^4 |\mathbf{u+v}|^2};\\
K^I_{55}(\mathbf{u},\mathbf{v}) &= \frac{(u^2-v^2)^2(\mathbf{u}\cdot \mathbf{v})^2(u^2v^2-(\mathbf{u}\cdot \mathbf{v})^2)}{4u^4 v^4 |\mathbf{u+v}|^2}. \\
\end{aligned}
\end{equation}
Note that they are symmetric in their arguments. The dependencies determining $I_{14}, I_{66}, I_{67}$ and $I_{77}$ are
\begin{equation}\label{depend}
\begin{aligned}
     28I_{12} - I_{22}+I_{23} -2\sqrt{6}(7I_{14}+5I_{24}-5I_{34}) &= 0; \\
     2I_{22}-2\sqrt{6}I_{24}+3I_{44}-18I_{66} &= 0; \\
     2I_{22} + 6I_{23}-5\sqrt{6}I_{24} - 3\sqrt{6}I_{34}+12I_{44}-72I_{67} &= 0; \\
     I_{22} + 6I_{23}+9I_{33}-4\sqrt{6}I_{24}-12\sqrt{6}I_{34} + 24I_{44} - 144I_{77} &=0.
\end{aligned}
\end{equation}
The (13)+(31) part of the power spectrum is given in terms of $J_n$, where the kernels are
\begin{equation}\label{jkernels}
\begin{aligned}
K^J_{1} &= \frac{10k^4p^4(k^2+p^2) - (21k^6 + 44k^4p^2+59k^2p^4)(\mathbf{k}\cdot \mathbf{p})^2 + 4(19k^2+7p^2)(\mathbf{k}\cdot \mathbf{p})^4}{42k^2p^4 |\mathbf{k+p}|^2 |\mathbf{k-p}|^2}; \\
K^J_2 &= \frac{-4k^2p^4(k^2+p^2)(17k^2+2p^2) + 8(18k^4p^2 + 25k^2p^4+3p^6)(\mathbf{k}\cdot \mathbf{p})^2 - 12(5k^2+13p^2)(\mathbf{k}\cdot \mathbf{p})^4}{21k^2p^4 |\mathbf{k+p}|^2 |\mathbf{k-p}|^2}; \\
K^J_3 &= \frac{1}{105k^4p^4 |\mathbf{k+p}|^2 |\mathbf{k-p}|^2}\bigg( -2k^4p^4(k^2+p^2)(29k^2+104p^2) \\
&+ k^2( 405k^4p^2+1042k^2p^4+705p^6)(\mathbf{k}\cdot \mathbf{p})^2 
+ 15(k^4-76k^2p^2-9p^4)(\mathbf{k}\cdot \mathbf{p})^4-360(\mathbf{k}\cdot \mathbf{p})^6 \bigg). \\
\end{aligned}
\end{equation}
This corrects the expressions for $I_{55}, J_3$ from \cite{Vlah20}. Note that $I_{55}$ is positive.

\section{Dependence on Line of Sight}
\label{App:D_L_S}
Throughout this paper, we have ignored the fact that modelling the measured projected ellipticities $\hat{\gamma}_{(1,2)}$ from Eq. \eqref{12} in principle requires additional contractions of $\Pi^{[n]}_{ij}$ with the line of sight\footnote{For a complete enumeration of selection effects in real space for scalar tracers, see for example \cite{Desjacques_2018_rs,Agarwal21}.}, as per the discussion in Section \ref{theory}. To quantify this, let us introduce the field 
\begin{equation}
    \tilde{g}_{ij}:= \text{TF}( \tilde{S}_{ij}), \quad \text{where } \quad \tilde{S}_{ij} := \frac{I_{ij}-\langle I_{ij}\rangle}{\frac{2}{3}\text{Tr}(I_{ij})}.
\end{equation}
Then, our measured estimator for $\gamma_{\pm 2}$ can be written as
\begin{equation}\label{selcor}
\begin{aligned}
     \hat{\gamma}_{\pm 2} = \frac{\frac{2}{3}\text{Tr}(I_{ij})\mathbf{M}_{ij}^{(\pm 2)*} \mathcal{P}_{ijkl}\tilde{g}_{kl}}{I_{11}+I_{22}}
     &= \frac{\frac{2}{3}\text{Tr}(I_{ij})\mathbf{M}_{kl}^{(\pm 2)*}\tilde{g}_{kl}}{\text{Tr}(I_{ij}) - \mathbf{\hat{n}}^i\mathbf{\hat{n}}^j I_{ij}} \\
    &= \frac{\frac{2}{3}\text{Tr}(I_{ij})\mathbf{M}_{kl}^{(\pm 2)*}\tilde{g}_{kl}}{\frac{2}{3}\text{Tr}(I_{ij}) - \mathbf{\hat{n}}^i\mathbf{\hat{n}}^j \text{TF}(I_{ij})} \\
    &= \mathbf{M}_{kl}^{(\pm 2)*}\tilde{g}_{kl} \bigg(\frac{1}{1-\mathbf{\hat{n}}^i\mathbf{\hat{n}}^j \tilde{g}_{ij}}\bigg) \\
    &= \mathbf{M}_{kl}^{(\pm 2)*}\tilde{g}_{kl}\bigg( 1 + \mathbf{\hat{n}}^i\mathbf{\hat{n}}^j \tilde{g}_{ij} + (\mathbf{\hat{n}}^i\mathbf{\hat{n}}^j \tilde{g}_{ij})^2 + \dots \bigg).
\end{aligned}
\end{equation}
Here we ignore the responsivity factor to avoid clutter. Thus, \textit{for this specific choice of normalisation of the shape field}, we can view the measured ellipticity fields as given by the theoretical ansatz from the main text (first term in the bottom line) plus some corrections due to the nonlinear projection dependence. This expression enables us to perform a \textit{self-consistency} check of our approach of neglecting these spurious dependencies: we can compute the correction by plugging in the same expansion for $\tilde{g}_{ij} $ we had before, and if the computed corrections to the multipoles are small, we can conclude \textit{a posteriori} that our approach was justified.  

We will not compute all relevant terms up to one-loop order here, but conclude with an illustrative example. Since the numerator of $\hat{\gamma}_{\pm 2}$ is first order in perturbations, this implies that the contribution of the leading correction (second term in the bottom line of Eq. \eqref{selcor}) to $P_{\delta E}(k,\mu)$ (or equivalently, $P_{\delta +}(k,\mu)$) is of the form 
\begin{equation}\label{sel}
\begin{aligned}
    P_{\delta E}(k,\mu) \supset \langle \delta^{(2)} \gamma_{+2}^{(2)} \rangle' &\supset (b_1^g)^2\mathbf{M}_{kl}^{(+2)*} \mathbf{\hat{n}}^i\mathbf{\hat{n}}^j \langle \delta^{(2)} K_{kl}^{(1)}K_{ij}^{(1)}\rangle'\\
    &= \frac{(b_1^g)^2}{2} \int \frac{d^3\mathbf{p}}{(2\pi)^3}\bigg\{\bigg(\frac{(\mathbf{m}^-\cdot \mathbf{p})^2}{p^2}\bigg)\bigg(\frac{(\mathbf{\hat{n}}\cdot (\mathbf{k-p}))^2}{|\mathbf{k-p}|^2}-\frac{1}{3}\bigg) + \bigg( \mathbf{p} \leftrightarrow \mathbf{k-p}\bigg)\bigg\} \\
    &\times F_2(\mathbf{p},\mathbf{k-p}) P_L(p) P_L(\mathbf{k-p}).
\end{aligned}
\end{equation}
By putting $\mathbf{\hat{k}}=(0,0,1)$ and $\mathbf{\hat{n}} = (\sqrt{1-\mu^2},0,\mu)$ and $\mathbf{p} = p(\cos \phi \sin \theta, \sin \phi \sin \theta, \cos \theta)$ for convenience (in contrast to the main body of this paper, but this is irrelevant), one can carry out the integration over $\phi$ and see that the $\mu-$dependence of this term will contain terms up to order $\mu^4$. A similar remark applies to higher order corrections (which do need to be taken into account at the one-loop level), as they also contain at least two powers of $\mathbf{\hat{n}}$. As a consequence, these corrections will spoil the relation $P_{\delta E}^{(0)}(k,\mu) + P_{\delta E}^{(2)}(k,\mu) = 0$ and it will give rise to a nontrivial hexadecapole $P_{\delta E}^{(4)}$. Specifically, the quantities $P_{\delta E}^{(0)} + P_{\delta E}^{(2)}$ and $P_{\delta E}^{(4)}$ directly probe these corrections. It was found that for this sample, $P_{\delta E}^{(0)} + P_{\delta E}^{(2)}$ and $P_{\delta E}^{(4)}$ are consistent with zero within $\sim 2\sigma$ over all scales included, so that we are justified in neglecting such corrections in the main analysis. This is confirmed by the direct computation of the loop integral in Eq. \eqref{sel}, which is at most $\mathcal{O}(10)$ on relevant scales. Thus, when multiplied by $(b_1^g)^2 \sim 6 \times 10^{-3}$ the resulting contribution falls almost entirely below the error bar of a single realization. However, upon inspecting the quantities $P_{\delta E}^{(0)} + P_{\delta E}^{(2)}$ and $P_{\delta E}^{(4)}$ for different choices of definition for the inertia tensor (see Appendix C of \cite{Kurita_2020}), it was found that some choices displayed a stronger deviation from the relation $P_{\delta E}^{(0)} + P_{\delta E}^{(2)} = P_{\delta E}^{(4)} = 0$. Thus, it seems unwarranted in general to ignore this additional line-of-sight dependence in the quasi-linear regime altogether. It is interesting to note that in this sense, the relative importance of selection effects is controlled by the size of the IA bias parameters; all contributions to multipoles from selection effects are proportional to higher powers of the IA parameters compared to the `isotropic' prediction. We plan to address all relevant contributions at one loop order explicitly in future work.

\bibliographystyle{JHEP}
\bibliography{mybibliography}
\end{document}